\documentclass[a4paper,11pt]{article}
\usepackage{jheppub}
\usepackage[T1]{fontenc}
\usepackage[utf8]{inputenc}
\usepackage{lipsum}
\usepackage{amsmath,amssymb,amsfonts,bbold}
\usepackage{braket}
\usepackage{natbib}
\usepackage[dvipsnames,svgnames,x11names]{xcolor}
\usepackage{graphicx,subfigure}
\usepackage{hyperref}
\usepackage{mathtools}
\usepackage[normalem]{ulem}
\newcommand{\be}{\begin{equation}}
\newcommand{\ee}{\end{equation}}
\newcommand{\bse}{\begin{subequations}}
\newcommand{\ese}{\end{subequations}}
\newcommand{\ba}{\begin{eqnarray}}
\newcommand{\ea}{\end{eqnarray}}
\newcommand{\bea}{\begin{eqnarray}}
\newcommand{\eea}{\end{eqnarray}}

\newcommand{\lb}{\left (}
\newcommand{\rb}{\right )}

\preprint{ CCTP-2025-14 \\ \hspace*{\fill}ITCP-2025-14}

\DeclareMathOperator{\sech}{sech}
\begin{document}


\title{On decoding the string from interfaces in 2d conformal field theories}

\author[1,2]{Avik Banerjee,}
\emailAdd{avik.banerjee@pwr.edu.pl}
\affiliation[1]{Institute of Theoretical Physics, Wroc\l{}aw  University  of  Science  and  Technology,  50-370  Wroc\l{}aw,  Poland}
\affiliation[2]{Crete Center for Theoretical Physics, Institute of Theoretical and Computational Physics,
Department of Physics, University of Crete, 70013 Heraklion, Greece}
\author[3]{Tanay Kibe,}
\emailAdd{tanay.kibe@wits.ac.za }
\affiliation[3]{National Institute for Theoretical and Computational Sciences,
School of Physics and Mandelstam Institute for Theoretical Physics,
University of the Witwatersrand, Wits, 2050, South Africa}
\author[4]{Ayan Mukhopadhyay,}
\emailAdd{ayan.mukhopadhyay@pucv.cl}
\affiliation[4]{Instituto de F\'{\i}sica, Pontificia Universidad Cat\'{o}lica de Valpara\'{\i}so,
Avenida Universidad 330, Valpara\'{\i}so, Chile.}
\author[5]{and Giuseppe Policastro}
\emailAdd{giuseppe.policastro@phys.ens.fr}
\affiliation[5]{Laboratoire de Physique de l'\'{E}cole Normale Supérieure, ENS, Universit\'{e} PSL, CNRS, Sorbonne Universit\'{e}, Universit\'{e} de Paris, F-75005 Paris, France}

\date{\today}

\abstract{General solutions of a gravitational junction between two copies of a three-dimensional Einstein manifold $\mathcal{M}$ correspond to the solutions of the non-linear Nambu-Goto equation for a  string in $\mathcal{M}$. 
We show that, for the junctions in three-dimensional anti-de Sitter spacetimes constituted by tensile strings, which are dual to interfaces between thermal states in conformal field theories, the solutions of the Nambu-Goto equation describing the junction correspond to wave-packets, which are perfectly reflected at the interface to future null infinity \textit{without shape distortion} when incident from past null infinity.
These wavepackets are realized by half-sided conformal transformations and affect the expectation value of the displacement operator. We further show that the entanglement entropy of an interval straddling the interface deciphers the stringy modes of the dual junction even in the tensionless limit. We also demonstrate that the strong sub-additivity of entanglement entropy is satisfied and is saturated for symmetric intervals generally.}

\maketitle

\section{Introduction} 
The holographic duality \cite{Maldacena:1997re,Gubser:1998bc,Witten:1998qj}, which states that quantum gravity with appropriate asymptotic boundary conditions can be reformulated as an ordinary quantum field theory living at the boundary of spacetime, is fundamental for the understanding of the emergence of spacetime  \cite{harlow2018tasi,Jahn:2021uqr,Chen:2021lnq,Kibe:2021gtw}. The holographic duality has also led to progress in our understanding of the quantum dynamics of strongly interacting quantum many-body systems by reformulating them as a classical theory of gravity in one higher dimension \cite{Hartnoll:2016apf}. Concrete ultraviolet complete examples of holographic duality have so far emerged only from string theory, a framework in which matter and gravity emerge together from the quantized vibrations of a fundamental string \cite{Green:2012oqa,Green:1987mn,Polchinski:1998rq,Polchinski:1998rr}. However, when the dual quantum field theory is strongly interacting, massive stringy modes decouple, leaving only classical gravity coupled to a few fields. Therefore, we can ask how the string plays a role in the holographic description of strongly interacting quantum theories. 

Junctions in three-dimensional anti-de Sitter (AdS$_3$) spacetimes correspond to interfaces in the dual two-dimensional strongly interacting conformal field theories (CFTs) with large central charges \cite{Karch:2000ct,Karch:2001cw,DeWolfe:2001pq,Bachas:2001vj}. The tension of the junction (string) characterizes the defect operator at the dual interface. 

It has been shown recently that the non-linear Nambu-Goto equation of the string emerges from the junction conditions of gravitational junctions in three-dimensional spacetimes \cite{Banerjee:2024sqq}. Therefore, understanding the holographic description of the stringy degrees of freedom of the gravitational junction has emerged as a question of fundamental importance. 

There has recently been progress in decoding the stringy degrees of freedom of the gravitational junction in terms of holographic quantum maps \cite{Chakraborty:2025dmc}. In this paper, we shall restrict ourselves to the simpler case in which the holographic interface is formed by gluing two identical CFTs each of which is in the same background state.  In \cite{Chakraborty:2025dmc}, it has been shown that each stringy vibration of the gravitational junction can be represented as a unique quantum map $\mathcal{H}_{\rm in}\rightarrow \mathcal{H}_{\rm out}$ that relates the incoming and outgoing excitations of the universal sector of the CFTs, which are scattered at the dual interface. Equivalently, each stringy vibration of the gravitational junction can be represented as a $\mathcal{H}_{\rm L}\rightarrow \mathcal{H}_{\rm R}$ quantum map relating the states of the two CFTs  connected by the interface. One remarkable feature is that the usual defect operator is generalized by a relative automorphism of the Virasoro algebras of the two CFTs, which is simply a relative conformal transformation at the linearized order in the incoming/outgoing fluxes. More precisely, the $\mathcal{H}_{\rm in}\rightarrow \mathcal{H}_{\rm out}$ quantum map is a composition of a universal energy scattering with a conformal transformation parametrized by the stringy excitation, which redistributes the energy in the in or out Hilbert space. Furthermore, the full quantum map is independent of the choice of the background state on both sides at the linear order in the incoming/outgoing fluxes.\footnote{Note that the quantum map is linear at all orders although the relationship between the incoming/outgoing energy fluxes are not. The automorphism of the Virasoro algebra  can be of the form $L_n\rightarrow S L_nS^{-1}$ with $S = \exp\left({\sum_{i}\alpha_i L_i +\sum_{ij}\tilde{\alpha}_{ij}L_i L_j +\cdots}\right)$. This automorphism, which is part of the full quantum map, can mix Virasoro generators non-linearly, e.g. $L_{-4}$ with $L_{-2}^2$. Therefore, the relationship between incoming and outgoing energy fluxes scattered at the interface need not be linear. See \cite{Chakraborty:2025dmc} for an explicit description of the generalized defect operator.}

The current paper has the motivation to understand the full non-linear structure of the quantum map corresponding to stringy vibrations of the gravitational junction, although not in the most general case. We focus on understanding the gravitational junction with two identical Ba\~nados-Teitelboim-Zanelli (BTZ) black brane geometries \cite{Banados:1992wn,Banados:1992gq}, representing thermal states of the identical CFTs across the dual interface, and with stringy vibrations corresponding to solutions of the non-linear Nambu-Goto equation in the BTZ background, as discussed in \cite{Banerjee:2024sqq}. We show that the full non-linear solutions of the gravitational junction can be translated to half-sided conformal transformations describing wavepackets which are reflected perfectly without distortion at the interface when the tension of the junction (string) is non-vanishing. 

In order to achieve a full non-linear understanding, we develop a method to resum the full perturbative expansion and gain a global understanding of the structure of the solution, adapting techniques developed in \cite{Banerjee:2023djb,Mitra:2024zfy} in other contexts. Furthermore, we show that the stringy vibrations of the gravitational junction are \textit{causally} borne out of initial conditions at past null infinity in the dual CFT, and the perfect reflection is manifestly causal.

We also address another important conceptual issue that arises in the limit of vanishing tension. Namely, the solutions of the non-linear equations describing the gravitational junction admit two rigid parameters that correspond to a rigid half-sided conformal transformation, which preserves the location of the interface and the thermal state. In the presence of these rigid parameters and the Nambu-Goto modes, the spacetime geometry is not smooth in the limit of vanishing tension. However, in this limit, the half-sided conformal transformation at the dual interface characterizing the junction remains rigid, so that the Nambu-Goto modes do not translate to any non-trivial reflected wavepackets at the dual interface. Therefore, it is natural to ask if there is a specific physical quantity in the interface CFT, which characterizes the Nambu-Goto modes of the junction. We show that the entanglement entropy of an interval straddling the interface indeed captures the Nambu-Goto modes even in the tensionless limit.

Given that quantum entanglement plays a fundamental role in explicit bulk reconstruction in holographic duality \cite{harlow2018tasi,Jahn:2021uqr,Chen:2021lnq,Kibe:2021gtw}, we take the first step towards understanding the reconstruction of the gravitational junction. We compute the entanglement entropy of a spacelike interval in the CFT in the presence of the rigid parameters and the Nambu-Goto modes. The entanglement entropy is a function of the temperature, the length of the interval, and the ratio of the lengths of the left to right parts of the interval straddling the interface. We find that the boundary entropy \cite{Affleck:1991tk,Cardy:2004hm,Afxonidis:2024gne,Azeyanagi:2007qj} is a function of the ratio of the left to right parts of the entangling interval only, i.e., it is unaffected by the temperature, rigid parameters, and the Nambu-Goto modes, which can be decoded, in the perturbative expansion, from the terms in the entanglement entropy that are quadratic in the length of the CFT interval. 
We also find that the strong sub-additivity (SSA) of the holographic entanglement entropy is always satisfied in the perturbative expansion in generic configurations, confirming that causality is indeed satisfied in the dual interface CFT. Furthermore, the SSA is always saturated in appropriate symmetric configurations, even in the presence of the stringy vibrations of the gravitational junction.

The plan of the paper is as follows. In Sec.~\ref{Sec:gravjunc}, we review the perturbative construction of the general solutions of the gravitational junction in \cite{Banerjee:2024sqq}, with two identical BTZ black holes on both sides and non-trivial excitations of the junction corresponding to the non-linear solutions of the Nambu-Goto equation in the BTZ background. In Sec.~\ref{Sec:holoint}, we holographically decode the stringy vibrations of the gravitational junction in terms of half-sided conformal transformations to all orders in the perturbative expansion. Furthermore, we resum the perturbative expansion and show that the half-sided conformal transformation can be expressed in terms of perfect reflection of wavepackets incident at the interface from one side. We show that the excitation of the interface corresponding to a stringy vibration of the junction originates causally from initial conditions at past null infinity. In Sec.~\ref{Sec:entanglement}, we study the holographic entanglement entropy of an arbitrary interval straddling the interface and find out how it encapsulates the stringy vibrations of the junction in the presence of the rigid parameters even in the tensionless limit. We also show that the strong sub-additivity of the holographic entanglement entropy is generically satisfied. In Sec.~\ref{Sec:conclusion}, we conclude with discussions on future directions.

\section{Gravitational junction}\label{Sec:gravjunc} 
Consider two copies $\mathcal{M}_{1,2}$ of a three manifold $\mathcal{M}$, each of which is split into two parts by co-dimension one hypersurfaces $\Sigma_{1,2}$, respectively. 
Note that the left ($L$) and right ($R$) parts are unambiguously specified, as $\mathcal{M}_{1,2}$ both inherit the coordinate charts of $\mathcal{M}$. A gravitational junction $\Sigma$ involves the union of one of the two fragments of $\mathcal{M}_1$ with one of the two fragments of $\mathcal{M}_2$, which we denote as $\mathcal{M}_{i\alpha_i}$ with $i = 1,2$ and the corresponding $\alpha_i = L,R$.
This union is formed by the identification of points of $\Sigma_1$ and $\Sigma_2$. Therefore, $\Sigma_{1,2}$ can be considered as the images of $\Sigma$ in $\mathcal{M}_{1,2}$, respectively.  The union of the fragments at the junction $\Sigma$ produces the full spacetime $\widetilde{\mathcal{M}}$. Identification of the points of $\Sigma_1$ and $\Sigma_2$, and their embeddings in $\mathcal{M}_1$ and $\mathcal{M}_2$, respectively, should satisfy the gravitational junction conditions.

It is useful to fix the freedom of choice of coordinates of the junction $\Sigma$ as follows. Let $t$ be the time coordinate, and let $z$ and $x$ be the spatial coordinates of $\mathcal{M}$, and so $\mathcal{M}_i$ have the coordinates $t_i$, $z_i$ and $x_i$. We can identify $x_i$ to be the coordinate transverse to $\Sigma_i$ and write the embeddings of $\Sigma_i$ in $\mathcal{M}_i$ as
\begin{equation}
    \Sigma_i: \,\, x_i = f_i(t_i, z_i).
\end{equation}
To form the junction $\Sigma$, we need to identify each point $P_1$ on $\Sigma_1$ with a unique point $P_2$ on $\Sigma_2$. The worldsheet coordinates $\tau$ and $\sigma$ of $\Sigma$ can be defined as
\begin{equation}
    \tau(P) =\frac{t_1(P_1) +t_2(P_2)}{2}, \,\,\sigma(P) =\frac{z_1(P_1) +z_2(P_2)}{2},
\end{equation}
i.e., as the average time and longitudinal embedding coordinates of $\Sigma_1$ and $\Sigma_2$, respectively. The junction is therefore specified by four variables, namely
\begin{equation}
    \tau_d = \frac{t_2 - t_1}{2},\,\, \sigma_d = \frac{z_2 - z_1}{2},\,\,x_{s,d} = \frac{f_2 \pm f_1}{2},
\end{equation}
all of which are functions of the worldsheet coordinates.

The junction conditions \cite{Israel:1966rt} can be derived from the action 
\begin{align}\label{Eq:bulk-action}
    S=\frac{1}{16\pi G_N}\int_{\mathcal{\widetilde{\mathcal{M}}}}d^3x \sqrt{-g}(R - 2\Lambda)+T_0\int_{\Sigma}{\rm d}\tau{\rm d}\sigma\,\sqrt{-\gamma}+{\rm GHY \ terms},
\end{align}
with the bulk metric $g$ being the only degree of freedom with respect to which it should be extremized. The Gibbons-Hawking-York (GHY) terms correspond to the boundaries of $\mathcal{M}_{i\alpha_i}$ (boundaries of $\mathcal{M}_{i\alpha_i}$ other than $\Sigma_i$ are not relevant for our analysis). $T_0$ is the tension of the string constituting the junction. The action \eqref{Eq:bulk-action} is defined with the assumption that the induced metrics $\gamma_i$ on $\Sigma_i$ are identical at each point $P(\tau,\sigma)$ on $\Sigma$, thus defining the worldsheet metric via
\begin{align}\label{Eq:hcont}
&\gamma_{\mu\nu}(\tau, \sigma) := \gamma_{1,\mu\nu}(\tau,\sigma) = \gamma_{2,\mu\nu}(\tau,\sigma).
\end{align}
Variation of the action \eqref{Eq:bulk-action} away from $\Sigma$ implies that
\begin{equation}
    R_{MN} -\frac{1}{2} R g_{MN} + \Lambda g_{MN} =0.
\end{equation}
The variation at the junction at $\Sigma$ yields
\begin{equation}\label{Eq:Kdisc}
    \sum_{i=1}^2 (-1)^{s(\alpha_i)}\left(K_{i,\mu\nu} + K_i\,\gamma_{\mu\nu}\right) = 8\pi G_N T_0 \gamma_{\mu\nu},
\end{equation}
with $s(\alpha_i) = 0$ if $\alpha_i = L$ and $s(\alpha_i) = 1$ if $\alpha_i = R$. $K_{i,\mu\nu}$ is the extrinsic curvature of $\Sigma_i$ in $\mathcal{M}_{i\alpha_i}$ and $K_i = \gamma^{\mu\nu}K_{i,\mu\nu}$. The symmetry of the action under bulk diffeomorphisms implies that the \textit{total} Brown-York tensor of the junction, which is the left hand side of \eqref{Eq:Kdisc}, is conserved. Therefore, we obtain only one independent equation from \eqref{Eq:Kdisc}. Together with \eqref{Eq:hcont}, the junction conditions give four independent equations, which determine the four variables $\tau_d$, $\sigma_d$, $x_s$ and $x_d$.

It has been shown that the general solutions of the junction conditions in three-dimensional gravity \cite{Banerjee:2024sqq} are in one-to-one correspondence with the solutions of the non-linear Nambu-Goto equation in $\mathcal{M}$, up to six rigid parameters which are related to worldsheet and spacetime isometries (see below).  For a junction gluing $\mathcal{M}_{1L}$ and $\mathcal{M}_{2R}$, the following holds generally.
\begin{enumerate}
    \item The hypersurface $$\Sigma_{NG}: t=\tau, \,\, z=\sigma,\,\, x = x_s(\tau,\sigma)$$in $\mathcal{M}$ (whose embedding coordinates are the average of those of $\Sigma_1$ and $\Sigma_2$) corresponds to a solution of the non-linear Nambu-Goto equations in $\mathcal{M}$ when the tension $T_0$ and the rigid parameters vanish. 
    \item Also, $x_s$ is the \textit{only} degree of freedom, since $x_d$, $\tau_d$ and $\sigma_d$ are completely determined as functions of $\tau$, $\sigma$, the tension and the rigid parameters given any choice of the solution of the Nambu-Goto equation in $\mathcal{M}$ corresponding to $x_s$.
\end{enumerate}
For a junction gluing $\mathcal{M}_{1R}$ and $\mathcal{M}_{2R}$, the roles of $x_s$ and $x_d$ are reversed. Without loss of generality, we will consider junctions gluing $\mathcal{M}_{1L}$ and $\mathcal{M}_{2R}$.

Here, we are interested in junctions in locally AdS$_3$ spacetimes that are dual to interfaces in the CFT. For interfaces in thermal backgrounds, $\mathcal{M}$ should be the BTZ black brane \cite{Banados:1992wn,Banados:1992gq} solution given by
\begin{equation}\label{Eq:BTZ}
    {\rm d}s^2 = \frac{ {\rm d}z^2}{\frac{z^2}{\ell^2}-4 \mu^2\ell^2}-\lb\frac{z^2}{\ell^2}-4\mu^2\ell^2\rb  {\rm d}t^2+\frac{z^2}{\ell^2}  {\rm d}x^2.
\end{equation}
Above, $\ell$ is the AdS$_3$ radius given by $\Lambda = -1/\ell^2$ and $4 \mu^2$ is the mass of the BTZ black brane. The dual interface lives at the boundary $z=\infty$. The Dirichlet boundary condition $x_1 =x_2 =0$ should also be imposed at the boundary, which corresponds to $\sigma =\infty$ on the worldsheet, so that the dual interface is positioned at $x=0$. This implies that
\begin{equation}\label{Eq:DBC}
    \lim_{\sigma\rightarrow \infty} x_s(\tau,\sigma) =\lim_{\sigma\rightarrow \infty} x_d(\tau,\sigma) =0.
\end{equation}
It is useful to define the variable $8\pi G_NT_0 = \lambda\epsilon$, and the rigid parameters as $\alpha_h\epsilon$ and $\gamma_h \epsilon$ (see below), with $\epsilon$ 
infinitesimal so that we can find the general solution perturbatively in $\epsilon$. The zeroth order solution corresponds to $x_s = x_d = \tau_d= \sigma_d =0$. For example, a specific solution with the boundary conditions \eqref{Eq:DBC} is
\begin{align}\label{Eq:wstoBTZ2}
    &x_{d}(\tau,\sigma)=\epsilon\left(-\frac{\lambda \ell^3}{2\sigma} \right) -\epsilon^2\frac{ 32 A_0 \ell^4 \mu ^5 e^{-6 \mu  \tau } \left(\gamma_h \sigma +\alpha_h e^{2 \mu  \tau } \sqrt{\sigma ^2-4 \ell^4 \mu ^2}\right)}{\sigma  \left(\sigma ^2-4 \ell^4 \mu ^2\right)^{3/2}}+ \mathcal{O}(\epsilon^3),\nonumber\\
    &x_{s}(\tau,\sigma)= \epsilon\left(-\frac{8 \ell^4\mu^4A_0e^{-4 \mu \tau}}{\sigma(\sigma^2-4\mu^2\ell^4)}\right) + \mathcal{O}(\epsilon^3),\nonumber\\
    &\tau_d(\tau,\sigma) = \epsilon\left(\alpha_h+\frac{\sigma}{2\sqrt{\sigma^2-4 \mu^2\ell^4}}\gamma_h e^{-2 \mu\tau}\right)+\epsilon^2\frac{ A_0 \lambda  \mu  e^{-4 \mu  \tau } \left(4 \mu ^2 \ell^4+\sigma ^2\right)}{4\ell \left( \sigma ^2-4\mu^2 \ell^4 \right)}+ \mathcal{O}(\epsilon^3),\nonumber\\
    &\sigma_d(\tau,\sigma) = \epsilon\left(\mu\sqrt{\sigma^2-4 \mu^2\ell^4}\gamma_h e^{-2 \mu\tau}\right)+\epsilon^2\frac{A_0 \lambda  \mu ^2 e^{-4 \mu  \tau } \left(4 \mu ^2 \ell^4+\sigma ^2\right)}{\ell \sigma }+ \mathcal{O}(\epsilon^3).
\end{align}
In the above solution, the leading term in $x_s$ is the leading quasi-normal mode, which is a solution of the linearized Nambu-Goto equation (about the zeroth order solution $x_s =0$) for a worldsheet in the BTZ background \eqref{Eq:BTZ} with the Dirichlet boundary condition \eqref{Eq:DBC} at the boundary of AdS and ingoing boundary condition at the worldsheet horizon. The amplitude of this leading quasi-normal mode is denoted by $A_0$. Note that the non-linear corrections from the Nambu-Goto equation appear at $\mathcal{O}(\epsilon^3)$ \cite{Banerjee:2024sqq}.  The rigid parameters $\alpha_h$ and $\gamma_h$ correspond to isometries that preserve the leading order (locally AdS$_2$) worldsheet metric.  One of such isometries which grows with $\tau$ as $e^{2\mu\tau}$ does not appear above as a perturbative solution does not exist if it is included. Three other rigid parameters related to relative spacetime isometries, which do not preserve $\Sigma_1$ (or $\Sigma_2$) but keep its extrinsic curvature invariant, can appear in general solutions but are ruled out by the Dirichlet boundary condition \eqref{Eq:DBC}. As claimed above, we notice from \eqref{Eq:wstoBTZ2} that $x_d$, $\tau_d$ and $\sigma_d$ are fully determined by $\lambda$, the two rigid parameters $\alpha_h$ and $\gamma_h$ and the choice of the solution of the Nambu-Goto equation corresponding to $x_s$. 

More generally, solutions of the Nambu-Goto equation in the BTZ background \eqref{Eq:BTZ}, which are ingoing at the worldsheet horizon, decay as $e^{-2n\mu\tau}$ at large $\tau$ with $n =2,3,\cdots$, and this follows from the behaviour of the quasi-normal modes. It follows that unless we reorganize the perturbative expansion, it is valid only at large $\tau$, since the quasi-normal modes grow as $\tau\rightarrow-\infty$. We will address this in Sec.~\ref{Sec:resum}.

\section{Holographic interpretation} \label{Sec:holoint}
Using the holographic dictionary, we can interpret the general gravitational junction as an interface between two semi-infinite wires, both described by a strongly interacting CFT with large central charge $c$ living at the boundary ($z=\infty$) \cite{Karch:2000ct,Karch:2001cw,DeWolfe:2001pq,Bachas:2001vj}. The coordinates for the first wire are $(t_1,x_1)$, with $x_1\leq 0$, while $(t_2,x_2)$ are the coordinates of the second wire, with $x_2\geq 0$, and the interface is located at $x_1=x_2=0$. 

The key feature of the general solutions relevant for the holographic interpretation of the stringy modes of the junction is that 
\begin{equation}
    \lim_{\sigma\rightarrow\infty}\tau_d(\tau,\sigma)=\mathbb{t}_d(\tau)
\end{equation}
is non-vanishing in the presence of quasi-normal modes, even when $\alpha_h =\gamma_h =0$, but with $\lambda\neq 0$, as evident from \eqref{Eq:wstoBTZ2}. A non-static solution of the junction conditions will necessarily feature a non-vanishing and non-constant $\mathbb{t}_d(\tau)$, implying relative time reparametrization at the dual interface, since $2 \mathbb{t}_d =t_2 - t_1$ at the boundary. Furthermore, the solution of the Nambu-Goto equation corresponding to the dual junction is encoded in $\mathbb{t}_d$ when $\lambda\neq 0$. We note that this is a generalization of the setup considered in earlier works that use gravitational interfaces as bottom-up models of dual conformal interfaces \cite{Karch:2000ct,DeWolfe:2001pq,Takayanagi:2011zk,Bachas:2020yxv,Bachas:2021tnp,Bachas:2021fqo,Liu:2024oxg}, which, however, don't feature a relative time reparametrization across the CFT interface.

Let us denote the normalizable mode of $x_s$ (which decays as $\sigma^{-3}$ asymptotically) as $\mathcal{A}(\tau)$, via
\begin{equation}\label{Eq:NGA}
    \lim_{\sigma\rightarrow\infty}x_s(\tau,\sigma)\sigma^3 = \mathcal{A}(\tau).
\end{equation}
Note that $\mathcal{A}$ is determined fully by the solution of the Nambu-Goto equation corresponding to the junction and behaves as $\mathcal{A}(\tau) \sim\sum_{n=2}^\infty \mathcal{A}_n e^{-2n\mu \tau}$ at large $\tau$. By solving the junction conditions in the radial expansion, we readily obtain the general relation
\begin{equation}\label{Eq:Tpar}
    {\dddot{\mathbb{t}}_{d}} - 4\mu^2 \dot{\mathbb{t}}_{d} = \frac{3}{2\ell^5}\epsilon \lambda \mathcal{A} + \mathcal{O}(\epsilon^3).
\end{equation}
One can easily verify the validity of the above relation in the solution \eqref{Eq:wstoBTZ2}.  

Since $\mathbb{t}_{d}(\tau)$ is the boundary value of $\tau_d(\sigma,\tau)$, we should regard each junction with a specified profile of $\mathbb{t}_{d}(\tau)$, the time jump across the interface, as a distinct interface in the dual CFT following the tenets of the holographic dictionary. We note from \eqref{Eq:Tpar} that $\mathbb{t}_{d}(\tau)$ specifies the solution of the non-linear Nambu-Goto equation in the bulk uniquely via the normalizable mode $\mathcal{A}$. In what follows, we will show that $\mathbb{t}_{d}(\tau)$ can be understood as a half-sided automorphism (conformal transformation) of the Virasoro algebra at the interface after transforming to physical continuous coordinates in which experiments are performed. The homogeneous solutions of \eqref{Eq:Tpar} correspond to half-sided $SL(2,R)$ isometries which do not affect the dual energy-momentum tensor.

To proceed further, we can use the relations $t_1 (\tau) = \tau - \mathbb{t}_d(\tau)$ and $t_2 (\tau) = \tau + \mathbb{t}_d(\tau)$, which hold at the boundary, to rewrite the relative time reparametrization in the form
\begin{equation}\label{Eq:Tpar2}
    t_2 = \mathbb{h}(t_1),
\end{equation}
at the interface $x_1 = x_2=0$. We can undo this relative time reparametrization at the interface  by performing the following conformal transformation on the second wire, which involves the diffeomorphisms
\begin{align}\label{Eq:CT}
    &\widetilde{t}_2 =\frac{1}{2}(\mathbb{h}^{-1}(t_2+x_2)+\mathbb{h}^{-1}(t_2-x_2)),\nonumber\\
    &\widetilde{x}_2 =\frac{1}{2}(\mathbb{h}^{-1}(t_2+x_2)-\mathbb{h}^{-1}(t_2-x_2)),
\end{align}
and the associated Weyl transformation that brings the metric on the second wire back to the Minkowski form. Setting $x_2 =0$ above, it is easy to see that the conformal transformation preserves the interface at $\tilde{x}_2=0$, where $\tilde{t}_2 = t_1$. Thus, we obtain both continuous coordinates and a continuous metric as a result of the half-sided conformal transformation given by \eqref{Eq:CT}. This half-sided conformal transformation can be uplifted to a bulk diffeomorphism $\mathcal{D}: X^M(\sigma,\tau) \rightarrow \tilde{X}^M(\sigma, \tau)$, an improper bulk residual gauge transformation ($x^M$ are the bulk coordinates) in $\mathcal{M}_{2R}$ which reproduces both the diffeomorphism \eqref{Eq:CT} and the associated Weyl transformation at the boundary $z_2=\infty$ \cite{deHaro:2000vlm}. Furthermore, both the induced metric on $\Sigma_2$ and the extrinsic curvature of the embedding of $\Sigma_2$ in $\mathcal{M}_2$ transform as scalars under a bulk diffeomorphism, and thus remain invariant. Therefore, we obtain an equivalent solution of the junction conditions (given by $\tilde{X}^M(\sigma, \tau)$) corresponding to the dual interface but with continuous coordinates and metric at the boundary. 

The holographic dictionary and holographic renormalization \cite{Henningson:1998gx,Balasubramanian:1999re} imply that the non-vanishing components of the energy momentum tensor on both wires, prior to the improper bulk diffeomorphism $\mathcal{D}$, are those of a thermal state at a temperature $T= \mu/\pi$, and are given by 
\begin{equation}
    T_{(1,2),\pm\pm} = (\pi c/12)T^2,
\end{equation}
with $c = 3\ell/2 G_N$. The anomalous transformation of the energy-momentum tensor under the conformal transformation \eqref{Eq:CT}, which is also reproduced by the dual improper bulk diffeomorphism $\mathcal{D}$ via the holographic dictionary \cite{Henningson:1998gx,Balasubramanian:1999re,deHaro:2000vlm}, implies that the energy momentum tensor on the second wire takes the form
\begin{align}\label{Eq:TR}
    &\widetilde{T}_{(2),\pm\pm}(\tilde{x}^\pm_2)= \mathbb{h}'(\tilde{x}^\pm_2)^2 T_{(2),\pm\pm}(\mathbb{h}(\tilde{x}_2^\pm))-\frac{c}{24\pi}{\rm Sch}(\mathbb{h}(\tilde{x}_2^\pm),\tilde{x}_2^\pm)\nonumber\\
    &\qquad\qquad\quad =\frac{\pi c}{12} \mathbb{h}'(\tilde{x}_2^\pm)^2 T^2 -\frac{c}{24\pi}{\rm Sch}(\mathbb{h}(\tilde{x}_2^\pm),\tilde{x}_2^\pm),
\end{align}
with $x_2^\pm = t_2 \pm x_2$ and $\widetilde{x}_2^\pm = \widetilde{t}_2 \pm \widetilde{x}_2$. The energy-momentum tensor on the first wire is unchanged in these continuous coordinates, as $\widetilde{T}_{(1),\pm\pm}={T}_{(1),\pm\pm}$.
As $\mathbb{h}$ is determined by $\mathbb{t}_d$, which in turn is determined by the solution of the non-linear Nambu-Goto equation corresponding to the dual gravitational junction, clearly the stringy modes at the junction are encoded in the left and right moving $\widetilde{T}_{(2),\pm\pm}(\tilde{x}_2^\pm)$ on the second wire. In other words, \textit{each solution of the non-linear Nambu-Goto equation corresponding to the dual gravitational junction translates to a wavepacket living on the second wire, which is reflected perfectly, without distortion, to future null infinity when incident on the dual interface from past null infinity, with both wires being at the same background temperature.} Note that the reflection is without distortion simply because $T_{++}$ and $T_{--}$ are the same function of their respective arguments. 
The half-sided conformal transformation preserves the conformal interface condition as it acts on left and right movers in the same way (see the next subsection).

Note that in the holographic junction, the boundary metric on either side of the junction itself does not change as a result of the half-sided conformal transformation (and the corresponding diffeomorphism which uplifts it to the bulk) although the coordinates at the boundary along with the boundary metric become continuous through the interface. The latter has allowed us to interpret the gravitational junction using the standard holographic dictionary.

\subsection{Perfect one-sided reflection as a special case of the general quantum map}

The perfect reflection of the wavepackets on the second (right) wire, which correspond to the Nambu-Goto modes, is consistent with the recently derived quantum maps for conformal interfaces dual to gravitational junctions with stringy excitations \cite{Chakraborty:2025dmc}. The gravitational junction corresponding to the static solution of the Nambu-Goto equation realizes the universal energy scattering of a conformal defect operator \cite{Meineri:2019ycm,Bachas:2020yxv}. This scattering can also be rewritten as a universal quantum map $\tilde{\mathcal{S}}:\mathcal{H}_{1}\to\mathcal{H}_{2}$ from the (universal sector of the) Hilbert space of the wire on the left of the interface to the Hilbert space of the wire on the right. This general map preserves the conformal boundary condition. As shown in \cite{Chakraborty:2025dmc}, the above map $\tilde{\mathcal{S}}$ is modified to the form $\mathcal{C}\circ\tilde{\mathcal{S}}$ in the presence of Nambu-Goto modes on the dual gravitational junctions. Here, $\mathcal{C}$ is a conformal transformation on the right wire, which depends on the Nambu-Goto modes, and acts in the same way on the left and right movers. Explicitly, the left and right moving energy fluxes of the right wire ($\widetilde{T}_{(2),\pm\pm}$) can be written in terms of the left and right moving energy fluxes on the left wire ($\widetilde{T}_{(1),\pm\pm}$)  as \cite{Chakraborty:2025dmc}
\begin{align}\label{Eq:HLHR}
   \widetilde{T}_{(2),++}(\omega)&=\lb 1+\frac{\lambda \ell}{2}\rb \widetilde{T}_{(1),++}(\omega) -\frac{\lambda\ell}{2}\widetilde{T}_{(1),--}(\omega) + \mathfrak{a}(\omega), \nonumber\\
    \widetilde{T}_{(2),--}(\omega)&= \frac{\lambda\ell}{2} \widetilde{T}_{(1),++}(\omega) +\lb 1-\frac{\lambda\ell}{2}\rb\widetilde{T}_{(1),--}(\omega) +\mathfrak{a}(\omega),
\end{align}
where the fluxes have the frequency $\omega$ on the left/right wire and $\mathfrak{a}$ is a Nambu-Goto mode (proportional to $\lambda (4-\lambda^2)^{3/4}\omega^{3/2}\mathcal{A}$ in \eqref{Eq:NGA}) of the gravitational junction in pure AdS$_3$ with the same frequency. It is easy to check from \eqref{Eq:HLHR} that 
\begin{equation}
   \widetilde{T}_{(2),++}(\omega) -\widetilde{T}_{(2),--}(\omega) = \widetilde{T}_{(1),++}(\omega) -\widetilde{T}_{(1),--}(\omega)
\end{equation}
implying the preservation of the conformal boundary condition. For  the $\omega\to0$ limit to be  well-defined, $\mathfrak{a}(\omega)$ should vanish at least as $\omega^3$ in this limit \cite{Chakraborty:2025dmc}.
These relations have been shown to hold at the linear order in $\tilde{T}_\pm^{L,R}$ but to all orders in $\lambda\ell$ \cite{Chakraborty:2025dmc}. In absence of the Nambu-Goto modes, 
\[
\begin{pmatrix}
    \widetilde{T}_{(2),++}(\omega)\\
     \widetilde{T}_{(2),--}(\omega)
\end{pmatrix} = \tilde{\mathcal{S}}\cdot \begin{pmatrix}
    \widetilde{T}_{(1),++}(\omega)\\
     \widetilde{T}_{(1),--}(\omega)
\end{pmatrix}, \quad \tilde{\mathcal{S}} = \begin{pmatrix}
    1+\frac{\lambda\ell}{2}\, &-\frac{\lambda\ell}{2}\\
    \frac{\lambda\ell}{2}\, & 1-\frac{\lambda\ell}{2}
\end{pmatrix}.
\]
The terms in \eqref{Eq:HLHR} proportional to the Nambu-Goto mode $\mathfrak{a}$ can be recast as a conformal transformation $\mathcal{C}$ acting in the same way on the left and right movers, so that the full quantum map $\mathcal{H}_{1}\to\mathcal{H}_{2}$ is $\mathcal{C}\circ\tilde{\mathcal{S}}$. In \cite{Chakraborty:2025dmc}, it was also shown that the full quantum map does not depend on the identical background state of the two CFTs straddling the dual interface (i.e. the identical locally AdS$_3$ spacetimes on both sides of the junction). 

We observe that \eqref{Eq:HLHR} has a solution of the form
\begin{align}\label{Eq:perf-refl}
    &\widetilde{T}_{(1),--}(\omega)= \widetilde{T}_{(1),++}(\omega) = 0, \quad \widetilde{T}_{(2),--}(\omega) = \widetilde{T}_{(2),++}(\omega) = \mathfrak{a}(\omega),\quad \omega\neq 0,\nonumber\\
    &\widetilde{T}_{(2),++} =\widetilde{T}_{(1),++}=\widetilde{T}_{(2),--} =\widetilde{T}_{(1),--} \quad {\rm at}\quad \omega =0
\end{align}
after setting $\mathfrak{a}(\omega) = O(\omega^3)$ for small $\omega$ as required \cite{Chakraborty:2025dmc} for perturbative analysis. This implies that perfect reflection of inhomogeneous energy modes on the right wire with both wires at the same background temperature is a special solution of the full quantum map. This is achieved by the half-sided conformal transformation on the right wire.

{Generally, the $\mathcal{C}\circ\tilde{\mathcal{S}}:\mathcal{H}_{1}\to\mathcal{H}_{2}$ map can be restated as $\mathcal{H}_{\rm in}\to\mathcal{H}_{\rm out}: \widetilde{\mathcal{C}}\circ\mathcal{S}$  or $\mathcal{H}_{\rm in}\to\mathcal{H}_{\rm out}: \mathcal{S}\circ \widetilde{\mathcal{C}}$ map preserving the conformal boundary condition with the scattering map $\mathcal{S}$ which is determined by $\lambda$ given by}
\begin{equation*}
\begin{pmatrix}
    \widetilde{T}_{(1),--}(\omega)\\
     \widetilde{T}_{(2),++}(\omega)
\end{pmatrix} = {\mathcal{S}}\cdot \begin{pmatrix}
    \widetilde{T}_{(1),++}(\omega)\\
     \widetilde{T}_{(2),--}(\omega)
\end{pmatrix}, \quad {\mathcal{S}} = \begin{pmatrix}
    \frac{\lambda\ell}{2+\lambda\ell}\, &\frac{2}{2+\lambda\ell}\\
\frac{2}{2+\lambda\ell}\, & \frac{\lambda\ell}{2+\lambda\ell},
\end{pmatrix}
\end{equation*}
{and $\widetilde{\mathcal{C}}$, a conformal transformation which achieves energy redistribution among incoming/outgoing energy modes \cite{Chakraborty:2025dmc}. Explicitly, at the linearized level, the energy modes transform as}
\begin{equation*}
   \widetilde{T}_{++}(\omega) \rightarrow  \widetilde{T}_{++}(\omega) - \frac{2 \mathfrak{a}(\omega)}{2+\lambda\ell},\quad \widetilde{T}_{--}(\omega) \rightarrow  \widetilde{T}_{--}(\omega) + \frac{2 \mathfrak{a}(\omega)}{2+\lambda\ell}
\end{equation*}
{as a result of $\widetilde{\mathcal{C}}$. Furthermore, it was also shown in \cite{Chakraborty:2025dmc} that the full quantum map is independent of the background state chosen identically on both sides. Thus the full quantum map $\mathcal{H}_{\rm in}\to\mathcal{H}_{\rm out}: \widetilde{\mathcal{C}}\circ\mathcal{S}$ explicitly at the level of linearized gravitational perturbations is}
\begin{align}\label{Eq:Hinoutfull}
   \widetilde{T}_{(1),++}(\omega) =\frac{\lambda\ell}{2+\lambda\ell}  \widetilde{T}_{(1),--}(\omega) + \frac{2}{2+\lambda\ell} \widetilde{T}_{(2),++}(\omega)-\frac{2 \mathfrak{a}(\omega)}{2+\lambda\ell},\nonumber\\
     \widetilde{T}_{(2),--}(\omega) = \frac{2}{2+\lambda\ell} \widetilde{T}_{(1),--}(\omega) + \frac{\lambda\ell}{2+\lambda\ell} \widetilde{T}_{(2),++}(\omega)+ \frac{2 \mathfrak{a}(\omega)}{2+\lambda\ell}.
\end{align} 
{Clearly in the absence of the Nambu-Goto modes ($\mathfrak{a}$), the full quantum map $\mathcal{H}_{\rm in}\to\mathcal{H}_{\rm out}: \widetilde{\mathcal{C}}\circ\mathcal{S}$ simplifies as the conformal transformation $\mathcal{\mathcal{C}}$ is just the identity transformation in this case. This reproduces the scattering of the static junction studied in \cite{Bachas:2020yxv} (with vanishing time jump $t_d(\tau)$ at the interface implemented explicitly by the Dirichlet boundary condition). As shown in \cite{Chakraborty:2025dmc}, the full quantum map $\mathcal{H}_{\rm in}\to\mathcal{H}_{\rm out}: \widetilde{\mathcal{C}}\circ\mathcal{S}$ implies that the general holographic interface is a tunable (and thus non-universal) energy transmitter. The scattering matrix $\mathcal{S}$ reproduces the universal scattering of conformal interfaces derived in \cite{Meineri:2019ycm} with an appropriate identification of $\lambda = 2(c- c_{LR})/c_{LR}$, where $c_{LR}$ is the central charge governing the two-point function of the stress tensor across the interface corresponding to the junction with vanishing Nambu-Goto modes.}

{It is easy to check that \eqref{Eq:perf-refl} is also a solution of \eqref{Eq:Hinoutfull} after setting $\mathfrak{a}(\omega) = O(\omega^3)$ for small $\omega$ as required \cite{Chakraborty:2025dmc} for perturbative analysis. Once again this implies that perfect one-sided reflection with equal background temperatures on both sides of the interface is a special case of the general $\mathcal{H}_{\rm in}\to\mathcal{H}_{\rm out}$ quantum map as well for any non-vanishing $\lambda$.}

{In our discussion on decoding the gravitational junction with stringy vibrations in the BTZ background in terms of an interface between thermal wires at the same temperature, we have shown that the full quantum map can be realized in the form of a half-sided conformal transformation acting in the same way on the left and right movers implementing \textit{perfect one-sided reflection to all orders in the perturbative expansion}. This generalizes the results of \cite{Chakraborty:2025dmc} obtained in linearized perturbation theory. The energy-momentum tensor on the second wire is given by \eqref{Eq:TR} and the Nambu-Goto modes $A_0$ (in BTZ and not pure AdS$_3$) correspond to wavepackets on top of the thermal background. We comment on more general constructions in the concluding section \ref{Sec:conclusion}.}

\paragraph{Ward identities:} At the interface dual to the gravitational junction, the Ward identities are satisfied with a  non-trivial expectation value of the displacement operator. To see this, it is useful to adopt the continuous coordinates $(\tilde{t}, \tilde{x})$ which coincide with $(t_1,x_1)$ for $x_1 \leq 0$ and with $(\tilde{t}_2,\tilde{x}_2)$ for $x_2\geq 0$, and are smooth at $x_1=\tilde{x}_2=0$ as mentioned above. The non-vanishing components of the energy momentum tensor then take the form
\begin{equation}\label{Eq:Tfull}
   \widetilde{T}_{\pm\pm}(\tilde{t},\tilde{x}) = \theta(-\tilde{x}) \widetilde{T}_{(1),\pm\pm}(\tilde{x}^\pm)+\theta(\tilde{x})\widetilde{T}_{(2),\pm\pm}(\tilde{x}^\pm),
\end{equation}
with $\widetilde{T}_{(1),\pm\pm} = {T}_{(1),\pm\pm}= (\pi c/12)T^2$, $\widetilde{T}_{(2),\pm\pm}$ given by \eqref{Eq:TR} and $\theta(0) = 1/2$. The Ward identities take the form
\begin{align}\label{Eq:WI}
    &\partial_{\tilde{t}}{\widetilde{T}}^{\tilde{t}\tilde{t}}+\partial_{\tilde{x}}{\widetilde{T}}^{\tilde{x}\tilde{t}}=0,\nonumber\\
    &\partial_{\tilde{t}}\widetilde{T}^{\tilde{t}\tilde{x}}+\partial_{\tilde{x}}\widetilde{T}^{\tilde{x}\tilde{x}}=\delta(\tilde{x})q(\tilde{t}),
\end{align}
with the transient source $q(\tilde{t})$ located at the interface given by
\begin{align}
    q(\tilde{t})= \left({\mathbb{h}'(\tilde{t})}^2-1\right) \frac{\pi c}{6}T^2-\frac{c}{12\pi}{\rm Sch}(\mathbb{h}(\tilde{t}),\tilde{t}),
\end{align}
which is the expectation value of the displacement operator \footnote{The expectation value of the displacement operator measures the energy cost of a small displacement of the interface \cite{Bianchi:2015liz}.}. The source in the first Ward identity in \eqref{Eq:WI} vanishes due to the conformal interface condition (energy conservation). For the specific solution \eqref{Eq:wstoBTZ2}, we obtain
\begin{equation}\label{Eq:hex}
   \mathbb{h}^{-1}(x)= x - \epsilon \lb 2 \alpha_h+\gamma_h e^{-2\mu x}\rb- \epsilon^2 \lb\frac{\mu  e^{-4 \mu  x} \left(A_0 \lambda +2 \gamma_h^2 \ell+4 \alpha_h \gamma_h \ell e^{2 \mu  x}\right)}{2 \ell}\rb +\mathcal{O}(\epsilon^3).
\end{equation}

When $\lambda$ vanishes, the time-reparametrization equation \eqref{Eq:Tpar} implies that the conformal transformation \eqref{Eq:CT} reduces to an isometry which preserves the thermal state. In this case, the source $q(t)$ in the Ward identity \eqref{Eq:WI} vanishes. We set $\gamma_h$ to zero, as otherwise the conformal transformation given by $\mathbb{h}^{-1}(x)$ maps the spatial infinity at any finite time to a finite point (see Sec.~\ref{Sec:resum}) . In this case, the conformal transformation \eqref{Eq:CT} reduces to just a half-sided time translation when $\lambda =0$.

In order to gain a deeper understanding, it is useful to reorganize the perturbative expansion so that it is valid for all times $\tau$ at each order in $\epsilon$. This can be achieved by adopting the methodology of \cite{Mitra:2024zfy}, as described in Sec.~\ref{Sec:resum}. As we show in the following subsection, the excitation in the dual gravitational junction described by the Nambu-Goto equation is born out of pre-existing initial conditions at past null infinity giving rise to the wavepacket incident at the interface, and we also recover these at future null infinity from the reflected wavepacket. This characteristic is manifested in the bulk coordinates $\tilde{X}^M$ corresponding to the dual half-sided conformal transformation \eqref{Eq:CT}.

\subsection{Re-summation of the perturbative expansion}\label{Sec:resum}
 In order to understand the global structure of the conformal transformation, it is necessary to go beyond large time and extend our analysis to all times by reorganizing the perturbative expansion. Firstly, in the absence of the quasi-normal modes, the rigid parameters $\alpha_h$ and $\gamma_h$ simply yield isometries of the worldsheet metric which translate to $\mathbb{h}^{-1}(x)$ in \eqref{Eq:CT} being an $SL(2,R)$ transformation which preserves the thermal state. Explicitly, in this case
\begin{equation}\label{Eq:his}
    \mathbb{h}^{-1}(x) = -\frac{1}{2\mu}\ln\lb\frac{a +b\,e^{-2\mu x}}{c+d\,e^{-2\mu x}} \rb,
\end{equation}
with $bc-ad =1$, and $b$ and $c$ non-vanishing. Generically, conformal transformations \eqref{Eq:CT} with $\mathbb{h}^{-1}(x)$ of the above form map the spatial infinity at any fixed time to a finite point. To avoid this we set $\gamma_h =0$, which reduces $\mathbb{h}^{-1}(x)$ to a translation of $x$ (with $a = d=0$). Therefore, \eqref{Eq:CT} reduces to a time translation in the absence of quasi-normal mode excitations at the junction. 

Taking advantage of the time reversal symmetry of the BTZ background \eqref{Eq:BTZ}, we note that at early time $\tau\rightarrow -\infty$ we can obtain a perturbative expansion simply by reversing the sign of $\tau$, as for instance in the solution \eqref{Eq:wstoBTZ2}. Such solutions are constituted of worldsheet modes which are outgoing at the worldsheet horizon at early time. However, physical solutions should not be outgoing at the worldsheet horizon at \textit{any} time. Adopting the methodology from \cite{Mitra:2024zfy}, developed in the context of understanding holographic Gubser flow (with results agreeing with exact solutions obtained numerically), we can achieve the construction of physical solutions at all times by reorganizing the perturbative expansion as follows. We require that at any order in $\epsilon$, the variables $\tau_d$, $\sigma_d$, $x_s$ and $x_d$ satisfy the following expansion in $\tau$
\begin{equation}\label{Eq:pert-rog}
    f(\tau,\sigma,\kappa) =w(\sigma)+ \sum_{n=2}^\infty \frac{u_n(\sigma)}{\cosh^n(2\mu \tau)} +\sum_{n=2}^\infty \frac{v_n(\sigma)}{(\kappa\cosh(2\mu \tau) +\sinh(2\mu \tau))^n},
\end{equation}
with $\kappa >1$ an arbitrary constant ($\kappa\cosh(2\mu \tau) +\sinh(2\mu \tau)>0$ for all $\tau$ if $\kappa >1$). We readily note that we recover the quasi-normal mode like expansion (i.e. expansion in powers of $e^{-2\mu\tau}$) at large $\tau$ but at the cost of doubling the number of coefficients. At early time $\tau\rightarrow-\infty$ we obtain the outgoing mode like expansion (i.e. expansion in powers of $e^{2\mu\tau}$). Requiring that the coefficients of $e^{2n\mu\tau}$ vanish at $\tau\rightarrow-\infty$ for $n\geq 2$, we can relate $v_n(\sigma)$ with $u_n(\sigma)$ so that \textit{both} of them are determined by the choice of coefficients of the late time expansion in powers of $e^{-2\mu\tau}$. Remarkably, the expansion \eqref{Eq:pert-rog} typically can be resummed for all $\tau$ although it depends on the choice of $\kappa$ \cite{Mitra:2024zfy} (an explicit example is in the next subsection). Therefore, \textit{the choice of the coefficients of the quasi-normal mode expansion of the solution of the Nambu-Goto equation determines the global solution of the junction conditions for any given $\kappa$}. More generally, one can consider the resummation
\begin{equation}\label{Eq:gennen}
  F(\tau,\sigma) = \int_1^\infty d\kappa \, \,c(\kappa)f(\tau,\sigma,\kappa),
\end{equation}
with a choice of weights $c(\kappa)$ satisfying $\int_1^\infty  d\kappa \,\,c(\kappa) = 1$, determining the global solution of the junction conditions together with the coefficients of the quasi-normal mode expansion of the solution of Nambu-Goto equation. As should be clear from the explicit example discussed in the next subsection, $f(\tau, \sigma,\kappa)$ coincides with the quasi-normal mode expansion for $\tau>0$ for all $\kappa >1$, but for $\tau <0$ behaves differently for different choices of $\kappa$; as a result \eqref{Eq:gennen} captures more general initial time behaviour.

We also note that by setting $\gamma_h =0$ we rule out any conformal transformation given by $\mathbb{h}^{-1}(x)$, which has a non-vanishing coefficient of $e^{-2\mu x}$ at large $x$ at a sub-leading order (irrespective of whether it preserves the thermal state or not). This implies that the conformal transformation given by
\begin{equation}
    \mathbb{h}^{-1}(x) = \frac{1}{2\tilde{\mu}}\ln\lb\frac{a +b\,e^{-2\mu x}}{c+d\,e^{-2\mu x}} \rb
\end{equation}
which changes the temperature of the second wire from $\mu/\pi$ to $\tilde{\mu}/\pi$ is also ruled out for any value of $a$, $b$, $c$ and $d$. 

It also follows that $\mathbb{h}^{-1}(x)$ determining the conformal transformation takes the form 
\begin{equation}\label{Eq:pert-rog-2}
    \mathbb{h}^{-1}(x,\kappa) =x+ h_w+ \sum_{n=2}^\infty \frac{h_{u_n}}{\cosh^n(2\mu x)}+\sum_{n=2}^\infty \frac{h_{v_n}}{(\kappa\cosh(2\mu x) +\sinh(2\mu x))^n}
\end{equation}
with $h_w$, $h_{u_n}$ and $h_{v_n}$ determined by the solution of the junction conditions. At $x\rightarrow-\infty$, the coefficients of the expansion in powers of $e^{2\mu x}$ should vanish.

Such a conformal transformation necessarily creates wavepackets \eqref{Eq:TR} whose departure from the thermal form decays at spatial infinity at large time, and in the infinite past/future at any spatial point (as each term in the above series decays at large values of the \textit{modulus} of their arguments), but has non-trivial profiles (in addition to the thermal form) at past and future null infinity (where $x^+$ or $x^-$ remains finite). $\widetilde{T}_{++}$ and $\widetilde{T}_{--}$  have such non-trivial profiles at past and future null infinities, respectively.  We discuss an explicit example in the next subsection.

\subsubsection{An explicit example}

\begin{figure}
    \subfigure[]{\includegraphics[width=0.5\linewidth]{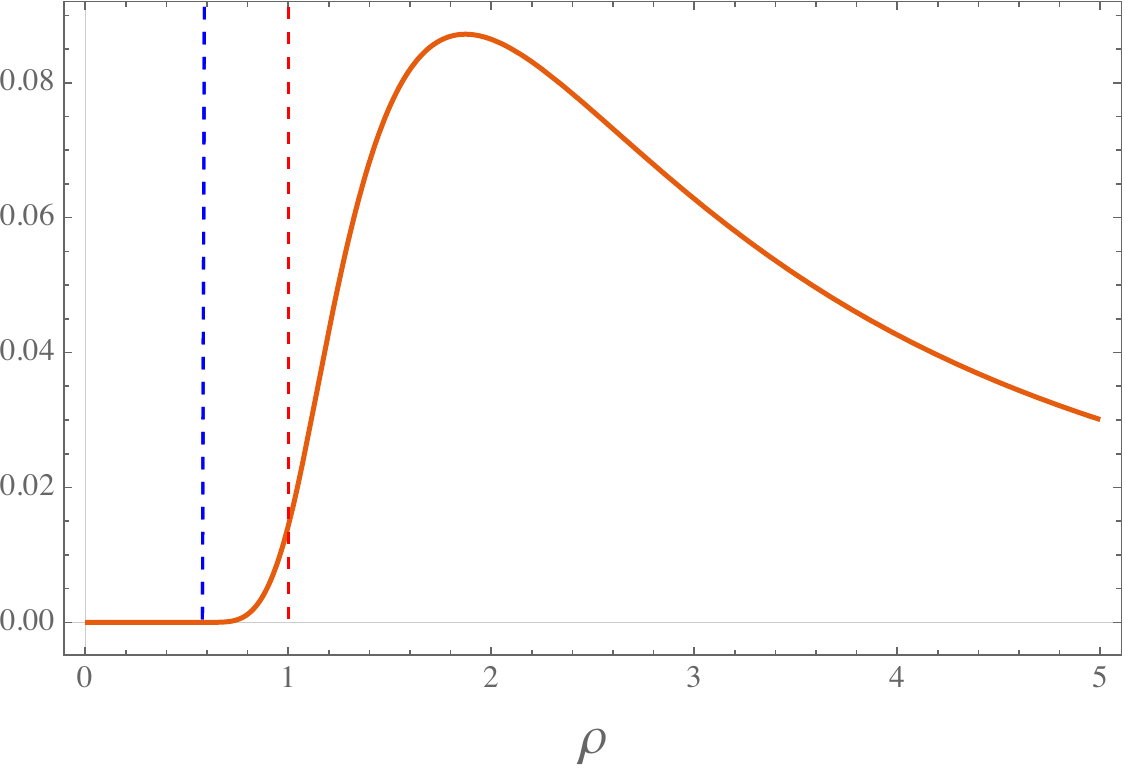}}
    \subfigure[]{\includegraphics[scale=0.29]{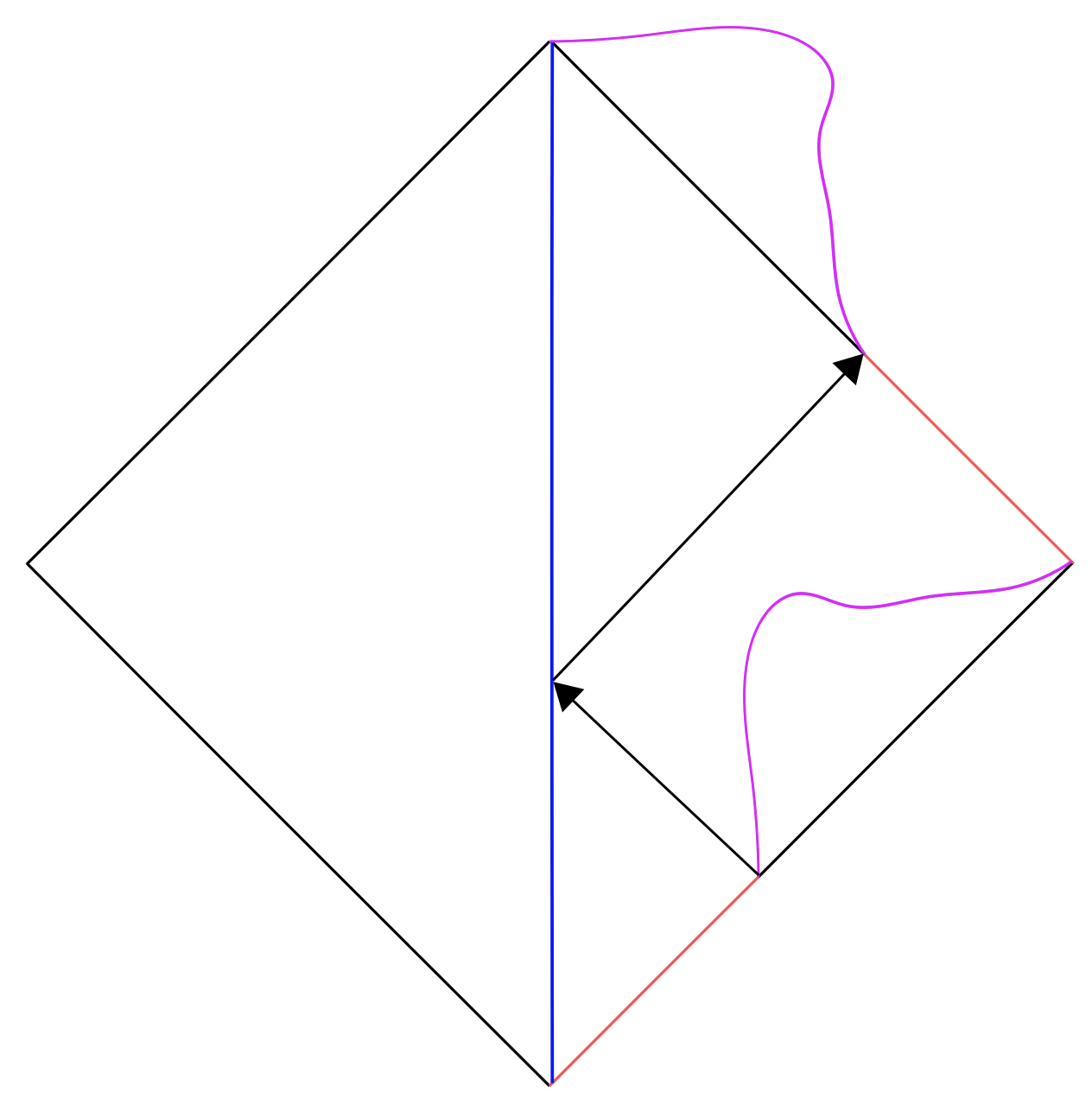}}
    \caption{(a) A plot of $\mathbb{h}^{-1}(\rho)$ with the parameters in \eqref{Eq:paramchoices}. The branch point at $\rho=1$ is the dashed red line and the branch point at $\rho=\sqrt{\frac{\kappa-1}{\kappa+1}}=\frac{1}{\sqrt3}$ is the blue dashed line. The function, its first, second and third derivatives are continuous. (b) Penrose diagram representing the perfect reflection in the dual interface CFT in two-dimensional Minkowski space. The solid black lines are the null infinities $\mathcal{I}^\pm$. The solid blue line is the interface. The departure, from the thermal state, of the state on $\mathcal{I}^-$ on the right CFT in the form of a wavepacket is indicated by the magenta function. The state departs from the thermal state for $\tilde{x}^+>-\frac{\ln3}{4\mu}$. The region of $\mathcal{I}^-$ where the state is thermal is shown in red. Similarly, on $\mathcal{I}^+$ we have the non-trivial state (wavepacket deformation) for $\tilde{x}^->-\frac{\ln3}{4\mu}$, and the thermal state (red) otherwise. As shown by the arrows, the non-trivial excitations on $\mathcal{I}^-$ hit the interface, are perfectly reflected, and then travel to $\mathcal{I}^+$. Causality of this process is manifest. }\label{fig:resumplot}
\end{figure}

The reorganized perturbative expansion readily allows us to construct explicit examples of wavepackets \eqref{Eq:TR}  which are reflected without distortion at the interface and which encode specific solutions of the Nambu-Goto equation corresponding to the dual junction. As for instance, we can extend the explicit solution \eqref{Eq:wstoBTZ2} by choosing the coefficients of the higher order quasi-normal modes so that the conformal transformation \eqref{Eq:CT} corresponding to the solution is given by
\begin{equation}
  \mathbb{h}^{-1}(x)= x+ \alpha_h +\sum_{n=2}^7 \alpha_{q_n} e^{-2n\mu x}
\end{equation}
at late time with no further sub-leading terms. The resummation given by \eqref{Eq:pert-rog-2} exactly converges for all values of the argument with $h_{v_n}$related to $h_{u_n}$ so that $\mathbb{h}^{-1 }(x)-x-h_w$ decays faster than any power of $e^{2\mu x}$ as $x\rightarrow-\infty$. Explicitly, with $\rho = e^{2\mu x}$, the resummation in this case \cite{Mitra:2024zfy} gives
\begin{align}\label{Eq:hexfull}
    \mathbb{h}^{-1 }(x) = \begin{cases}
        x+ \alpha_h + \sum_{n=2}^7 \alpha_{q_n} \rho^{-n} \,\,&{\rm for}\,\, x>0,\\
        x+\alpha_h + \sum_{n=2}^7 \alpha_{q_n} \frac{-(\kappa-1)^n + (\kappa+1)^n \rho^{2n}}{(-(\kappa-1)^n+(\kappa+1)^n)\rho^n} \, \,& {\rm for}\,\, \frac{1}{4\mu}\ln\lb\frac{\kappa -1}{\kappa+1} \rb\leq x \leq0, {\rm and}\\
        x+ \alpha_h\,\,&{\rm for}\,\, x <\frac{1}{4\mu}\ln\lb\frac{\kappa -1}{\kappa+1}\rb. \,\, 
    \end{cases}
\end{align}
Note that $\mathbb{h}^{-1 }(x)$ is continuous at $x=0$ (i.e. $\rho =1$) and also at $x = \frac{1}{4\mu}\ln\lb\frac{\kappa -1}{\kappa+1}\rb$ (i.e. $\rho =\sqrt{\frac{\kappa-1}{\kappa+1}}$)\footnote{Note that $\cosh(2\mu x)$ is minimum at $x=0$ whereas $\kappa\cosh(2\mu x) + \sinh(2\mu x)$ is minimum at $x = \frac{1}{4\mu}\ln\lb\frac{\kappa -1}{\kappa+1}\rb$ (i.e. $\rho =\sqrt{\frac{\kappa-1}{\kappa+1}}$). These give the branch points of $\mathbb{h}^{-1 }(x)$.}. However, to obtain smooth wavepackets given by \eqref{Eq:TR}, we need $\mathbb{h}^{-1 }(x)$ to have continuous first, second and third derivatives for all values of the argument. This can be readily achieved, as for instance by setting
\begin{equation}
    \kappa = 2,
\end{equation}
and
\begin{align}\label{Eq:paramchoices}
        &\alpha_{q_3} = \frac{561218-306612 \sqrt{3}}{81 \left(16
   \sqrt{3}-385\right)}\alpha_{q_2},\nonumber\\
   &\alpha_{q_4} = -\frac{2 \left(6120058
   \sqrt{3}-9722513\right)}{1327113}\alpha_{q_2},\nonumber\\
    &\alpha_{q_5} = \frac{40 \left(25333
   \sqrt{3}-26333\right)}{442371}\alpha_{q_2},\nonumber\\
   &\alpha_{q_6} =\frac{872187357-492482732
   \sqrt{3}}{107496153} \alpha_{q_2},\nonumber\\
   &\alpha_{q_7} =\frac{2 \left(2243122
   \sqrt{3}-4141367\right)}{1327113} \alpha_{q_2}.
\end{align}
See Fig.~\ref{fig:resumplot} (a) for a plot of $\mathbb{h}^{-1}(x)$, with the above choice of parameters.
The smooth wavepackets obtained from \eqref{Eq:TR} via the half-sided conformal transformation \eqref{Eq:hexfull} in this case have non-trivial departures from the thermal state for 
\begin{equation}
    \tilde{x}^+ >-\frac{1}{4\mu}\ln3 
\end{equation}
at past null infinity ($\tilde{x}^-\rightarrow -\infty$) and for 
\begin{equation}
    \tilde{x}^- >-\frac{1}{4\mu}\ln3 
\end{equation}
at future null infinity ($\tilde{x}^+\rightarrow \infty$). Also, the stringy modes are excited at the junction at $\tilde{t} =-\frac{1}{4\mu}\ln3 $ when the left moving wavepacket is incident from the past null infinity at the dual interface. This example illustrates that the stringy excitations of the junction are born out of the pre-existing initial conditions at past null infinity which are recovered back at future null infinity. Crucially, the full process is causal -- see Fig.~\ref{fig:resumplot} (b) for an illustration.

\section{Quantum entanglement deciphers the string}  \label{Sec:entanglement}
Quantum entanglement in conformal interfaces has been studied extensively in the literature \cite{Azeyanagi:2007qj,Sakai:2008tt,Brehm:2015lja,Gutperle:2015kmw,Gutperle:2015hcv,Wen:2017smb,Goto:2020per,Tang:2023chv,Afxonidis:2024gne,Afxonidis:2025jph,Jokela:2025qac}. Particularly, the entanglement entropy of two-dimensional holographic interface CFTs can be studied using the HRRT (Hubeny-Rangamani-Ryu-Takayanagi) prescription \cite{RT,HRT}, which states that the entanglement entropy of an interval in the dual CFT is given by 
\begin{equation}
    S=\frac{L_{\rm geo}}{4G_N},
\end{equation}
where $L_{\rm geo}$ is the length of the shortest bulk geodesic that connects the end-points of the boundary interval, and which is homologous to the boundary interval. The entropy is computed with a bulk radial cut-off that is identified with the ultraviolet regulator of the dual theory. We focus on the case when the interface is between two CFTs with the same central charge $c$, which is given by $c = 3\ell/2 G_N$ \cite{Brown:1986nw,Henningson:1998gx,Balasubramanian:1999re}.

In the presence of a defect (i.e. when $\lambda\neq 0$), the half-sided conformal transformation which encodes the stringy vibrations of the dual junction, cannot factorize. So we expect that the entanglement entropy of an interval straddling the dual holographic interface should reveal the Nambu-Goto solution corresponding to the dual gravitational junction. We can explicitly see this by computing the entanglement entropy via the HRRT prescription and adopting the methodology developed in \cite{Kibe:2021qjy,Banerjee:2022dgv,Kibe:2024icu}, which can be adopted in the presence of junctions in locally AdS$_3$ spacetimes. See Appendix~\ref{App:EE} for details on the method. Since the solution of the junction conditions has been evaluated perturbatively, the entanglement entropy should be obtained in the $\epsilon$ expansion. As the perturbative expansion breaks down at the worldsheet horizon, we should restrict ourselves to intervals of length $l$ for which $\mu l \ll 1$, and also expand the result in powers of $\mu l$. Moreover, the holographic entanglement entropy is computed with $\sigma_c = \ell^2\delta_c^{-1}$ as the radial cutoff, which implements the UV cutoff $\delta_c$ in the CFT. This implies that the boundary intervals should also satisfy $\mu \delta_c\ll \mu l$.

Furthermore, we can define the following dimensionless physical quantities using the AdS radius $\ell$ and $\mu$
\begin{equation}
    \ln g_0 = \frac{c}{6}\lambda \ell,\, \,A_\mu =\frac{c}{6} \frac{\mu^2 A_0}{\ell^2}, \,\,\alpha_\mu = \frac{c}{6}\alpha_h \mu.
\end{equation}
Above, $g_0$ ($\lambda$) is related to the boundary entropy \cite{Affleck:1991tk,Cardy:2004hm,Azeyanagi:2007qj} of the interface and characterizes the defect operator. For an interval straddling the defect, a half-sided conformal transformation \eqref{Eq:CT} (valid for $\tilde{x}>0$) acting on the right end-point as $(t_2,x_2)\rightarrow (\tilde{t}_2,\tilde{x}_2)$ and holding the left endpoint $(t_1,x_1)$ fixed, transforms the entanglement entropy $S$ of the interval to $\tilde{S}$, as \cite{Calabrese:2004eu}
\begin{align}
    \tilde{S}(t_1,x_1;t_2,x_2) = S(t_1,x_1;\tilde{t}_2,\tilde{x}_2)-\frac{c}{12}\ln\lb {\mathbb{h}^{-1}}'(x_2^+){\mathbb{h}^{-1}}'(x_2^-)\rb.
\end{align}
This anomalous transformation is reproduced by the HRRT prescription. 

We are interested in the entanglement entropy of an equal time interval at time $t_0$ in the continuous $(\tilde{t},\tilde{x})$ coordinates, which are described in Sec.~\ref{Sec:holoint}. Therefore, we choose 
\begin{equation}
    t_1 = \tilde{t}_2 =t_0, \quad x_1 = -ly/(1+y),\quad \tilde{x}_2 = l/(1+y),
\end{equation}
with the interface located at $x_1 =\tilde{x}_2 =0$. Note that the length of the interval is $l$ and $y$ is the ratio of the length of the part of the interval on the left to that of the part on the right of the interface. For the solution \eqref{Eq:wstoBTZ2} of the dual junction with $\gamma_h =0$, the entanglement entropy of this interval with the anomalous part subtracted is given by
\begin{align}\label{Eq:EEsubtr}
  S &= \frac{c}{3} \ln \frac{l}{\delta_c}+\ln g_{\rm eff}(y)+
   \mu^2 l^2\Bigg( \frac{c}{18}+\epsilon  \lb\frac{\ln g_0 (y-1)^2\sqrt{y}}{3 (y+1)^3}\rb +\nonumber\\
    &+\epsilon^2\frac{2}{c}\Bigg( -\frac{(\ln g_0)^2  y (y-1)^2}{(1+y)^4}+\frac{384 A_\mu \alpha_\mu y e^{-4 \mu  t_0} \sqrt{y}  }{ (1+y)^3}-\frac{12 (\ln g_0) A_\mu e^{-4 \mu  t_0}\left(y^2-4y-1\right)}{(y+1)^3}\Bigg)+\\&  \qquad \qquad \mathcal{O}(\epsilon^3)\Bigg)+ \mathcal{O}(\mu^3l^3)\nonumber,
\end{align}

where
\begin{equation}\label{Eq:geff}
    \ln g_{\rm eff}(y)=\epsilon \lb\frac{2\sqrt{y} \ln g_0}{1+y}\rb+
    \epsilon^2\frac{6}{c} \lb\frac{(\ln g_0)^2 (y-1)^2}{4 (y+1)^2}\rb+
    \mathcal{O}(\epsilon^3)
\end{equation}
is the boundary entropy \footnote{The interface CFT is related to a boundary CFT via a folding trick \cite{Oshikawa:1996dj,Bachas:2001vj,Quella:2006de}}, which is determined only by $\ln g_0$ (i.e. $\lambda$) and which is independent of $l$ and the time $t_0$. The function $g_{\rm eff}$ depends only on the ratio $y$ \cite{Afxonidis:2024gne}, as expected from scale invariance. We note that the above is valid when the lengths of the left and right parts of the interval are larger than $\delta_c$, i.e. $y/(1+y) > \delta_c/l$ and $(1-y)/(1+y) > \delta_c/l$ \footnote{Entanglement entropies of intervals with one end-point on the defect can have modified effective central charges $c_{\rm eff}$ that multiply the leading log divergence (see for example \cite{Afxonidis:2025jph}). This effect cannot be seen by simply taking the $y\to0$ limit of our perturbative answer. We postpone an analysis of $c_{\rm eff}$ to a future publication.}. For an interval that is symmetric across the defect we have $y=1$ and we reproduce the boundary entropy obtained for a symmetric interval in \cite{Azeyanagi:2007qj}. As far as we are aware, \eqref{Eq:EEsubtr} is the first explicit computation of $g_{\rm eff}(y)$ in the setup considered in this paper and also in \cite{Azeyanagi:2007qj}. In general, the function $g_{\rm eff}$ is expected to be complicated, theory-dependent and not universal \cite{Chapman:2018bqj,Karch:2021qhd,Anous:2022wqh,Afxonidis:2024gne}. It is to be noted that the expansion in $\epsilon$ and $\mu l$ together constitute the general low-temperature expansion. Subtracting the vacuum contributions ensures that the $\epsilon$ and $\mu l$ expansion coefficients are cut-off independent.

Remarkably, $\alpha_h$ and the Nambu-Goto amplitude $A_0$ are visible only at order $\mu^2l^2$ in the $\mu l$ expansion. We note that when $\lambda=0$ and $\alpha_h=0$ in \eqref{Eq:EEsubtr} we recover the $\mu l$ series expansion of the entanglement entropy of a thermal state dual to a BTZ geometry, which is
\begin{equation}
    S_{\rm BTZ} =\frac{c}{3} \ln \frac{l}{\delta_c}+ \frac{c}{3} \log \lb\frac{\sinh(\mu l)}{\mu l}\rb \approx \frac{c}{3} \ln \frac{l}{\delta_c}+\frac{c}{18}\mu^2l^2+\mathcal{O}(\mu^3l^3).
\end{equation}
Thus, the departure of the entropy from the thermal form (given just by the continuous BTZ background) vanishes when both $\lambda$ and $\alpha_h$ are taken to zero, since in this case $x_d \rightarrow 0$, although $A_0\neq 0$ (see \eqref{Eq:wstoBTZ2} for an example), and therefore the gravitational junction reduces to a smooth spacetime. In this limit, the half-sided conformal transformation just becomes the identity. However, in the tensionless limit $\lambda=0$ there are non-vanishing corrections to the quadratic in $l$ terms of the entanglement entropy depending on $\alpha_h A_0$. Although the half-sided conformal transformation reduces just to time translation when $\lambda =0$, implying the absence of wavepackets that are reflected at the interface, the gravitational junction corresponds to a non-trivial spacetime (with $x_d\neq 0$) and the entanglement entropy still deciphers the Nambu-Goto modes. 

The first general lesson is thus that the Nambu-Goto solution corresponding to the gravitational junction can be deciphered from the quadratic in $l$ terms of an interval straddling the dual interface. Secondly, the limit $\lambda\rightarrow 0$ is subtle in the presence of a rigid half-sided time translation and wavepackets which are reflected without distortion at the dual interface in a holographic CFT with a large central charge. This subtlety is captured by the gravitational junction -- to obtain a non-smooth spacetime at vanishing $\lambda$, we need to first construct solutions of the junction conditions with $\lambda \neq 0$ and $\alpha_h\neq0$, and then take the limit $\lambda\rightarrow 0$. 

\begin{figure}
    \subfigure[]{\includegraphics[width=0.45\linewidth]{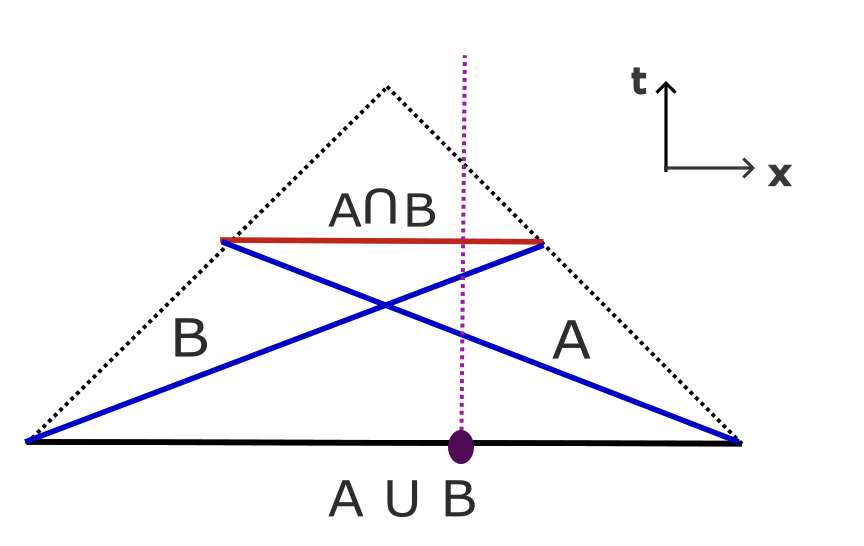}}
    \subfigure[]{\includegraphics[width=0.45\linewidth]{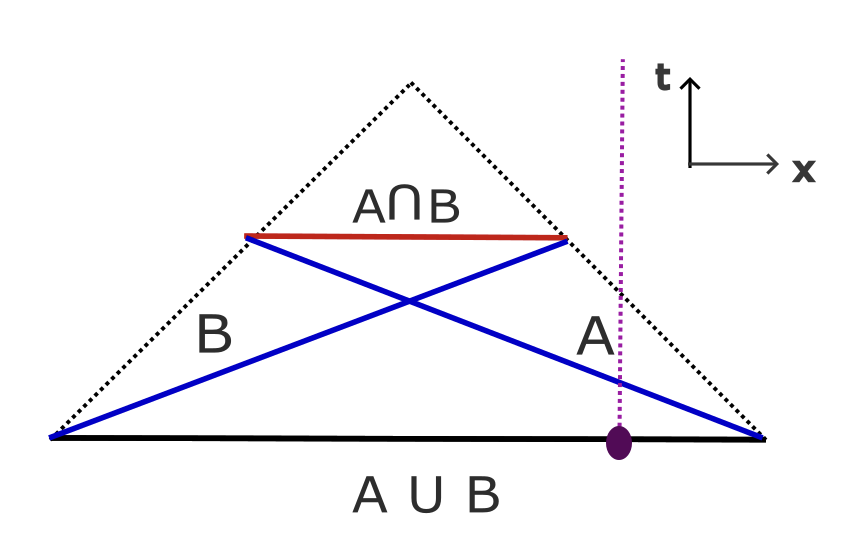}}
    \caption{The four regions $A \cup B$, $A$, $B$ and $A\cap B$ involved in the SSA are shown with (a) and (b) denoting two different configurations. The dotted black lines are the light cone directions. The purple line is the trajectory of the interface.}
    \label{Fig:SSA1}
\end{figure}

{Furthermore, the entanglement entropy reveals that the tensionless junction with Nambu-Goto modes is dual to a non-trivial interface in the presence of rigid parameter $\alpha$ (giving half-sided time translation). It can be seen from \eqref{Eq:Hinoutfull} that such a tensionless interface is purely transmitting (as the reflection coefficient vanishes and the transmission coefficient becomes unity when $\lambda =0$), and therefore topological as $\mathfrak{a}(\omega)$ (which are proportional to $\lambda \mathcal{A}(\omega)$, with $\mathcal{A}(\omega)$ being the amplitudes of the Nambu-Goto modes as mentioned before) also vanish in the tensionless limit. So, the tensionless junction with Nambu-Goto modes realizes a \textit{novel non-trivial topological interface} in the dual CFT in the presence of the rigid parameter $\alpha$.}

{It is important to understand the construction of such topological interfaces in the CFT explicitly. We recall that the Nambu-Goto modes do not lead to a non-trivial half-sided conformal transformation in the tensionless limit in the large $N$ limit. However, the large $N$ limit can be subtle. \footnote{It is of interest to see if the holographic junction with(out) Nambu-Goto modes and its non-trivial tensionless topological limit can be described better in terms of a relation between the commuting quantum KdV hierarchy of charges \cite{Bazhanov:1994ft} across the interface. The large $N$ limit can be subtle if not all the quantum KdV charges scale as $N^2$.} A detailed understanding of how the entanglement entropy can be reproduced directly in the CFT should enrich our understanding of semi-classical spacetime.}

\subsection{Strong sub-additivity}\label{Sec:SSA}
In a relativistic local quantum field theory, intervals with the same end points and the same domain of dependence have the same entanglement entropy. The satisfaction of strong sub-additivity of entanglement entropy can be used as a non-trivial check of the absence of causality violation in a relativistic theory and also to impose constraints on $g_{\rm eff}$ \cite{Azeyanagi:2007qj,Afxonidis:2024gne} \footnote{See \cite{Casini:2022bsu} for information theoretic constraints on the boundary entropy in general spacetime dimensions.} by considering the four intervals $A$, $B$, $A\cup B$ and $A\cap B$ with endpoints on a lightcone as shown in Fig.~\ref{Fig:SSA1}. The domains of dependence of $A\cup B$ and $A\cap B$ are the union and intersections of the domains of dependence of $A$ and $B$, respectively \cite{Casini:2006es}. Note also that the lightcone is not affected by the half-sided conformal transformation. 

We find that in both cases, whether the defect intersects $A$ and $B$ or only one of them, the strong sub-additivity of the entanglement entropy given by
\begin{equation}
    S(A) + S(B)-S(A\cup B)-S(A \cap B)\geq0
\end{equation}
is satisfied perturbatively. Note that the SSA is satisfied even when $\gamma_h \neq0$. See Appendix~\ref{App:SSA} for details.

Note that the anomalous contribution to the entanglement entropies coming from the half-sided conformal transformation cancel out on the left hand side of the above inequality, and as in the case of the equal time interval, the Nambu-Goto modes and $\alpha_h$ do not affect the non-anomalous parts of the entanglement entropy at the leading order in the $\mu l$ expansion. It is known \cite{Azeyanagi:2007qj} that the leading order $(\mu l)^0$ terms in holographic interface CFT satisfy the SSA and saturate it only in the symmetric case in which $A$ and $B$ intersect each other at the interface. Remarkably, we find that the SSA is saturated in the symmetric case even in the presence of $\alpha_h$ and the Nambu-Goto modes for any value of $\lambda$ and $\mu$ within the realm of the perturbative expansions (see Appendix~\ref{App:SSA}).  It would be interesting to understand this characteristic using CFT methods.

\section{Conclusions}\label{Sec:conclusion}
In this paper, we have shown that stringy degrees of freedom of a gravitational junction gluing two BTZ (locally AdS$_3$) spacetimes can be translated to wavepackets that can reflect without distortion at the dual interface joining two thermal wires at the same temperature. The transient sources appearing in the Ward identities for energy and momentum conservation at the interface, which encapsulate the stringy degrees of freedom of the dual junction, are born out of pre-existing initial conditions in the past null infinity that are recovered back at the future null infinity. We have also shown that the stringy modes of the junction can be deciphered from the terms in the entanglement entropy of intervals straddling the dual interface which are quadratic in the length of the interval, even when the tension of the dual junction vanishes.

{It is of primary importance to understand the junction explicitly, and particularly the non-trivial topological tensionless limit in the presence of the Nambu-Goto modes and the half-sided time translation, in the dual CFT. The reproduction of the non-trivial entanglement in the tensionless limit, which is obtained in the presence of the Nambu-Goto modes and the half-sided time translation, directly in the CFT should deeply enrich our understanding of holographic reconstruction \cite{harlow2018tasi,Jahn:2021uqr,Chen:2021lnq,Kibe:2021gtw} of semi-classical spacetimes.} 

{It would be fascinating to explore the gravitational junction with different temperatures on both sides of the junction. In the absence of the Nambu-Goto modes, it has been shown that the dual interface develops a steady state heat current in the presence of different temperatures on both sides \cite{Bachas:2021tnp}. It would be particularly interesting to explore whether the Nambu-Goto modes of the junction imply that the heat flow in the general dual interfaces can be manipulated as suggested by the linearized general quantum map studied in \cite{Chakraborty:2025dmc}.}

The results of the two-way gravitational junctions in locally AdS$_3$ spacetimes admit generalization to multi-way junctions \cite{Chakraborty:2025jtj} gluing $n >2$ locally AdS$_3$ spacetimes.\footnote{Multiway junctions in AdS$_2$ were studied earlier in \cite{Shen:2024dun,Shen:2024itl}. See also \cite{Liu:2025khw}.} As shown in  \cite{Chakraborty:2025jtj}, the general solutions of the $n$-way gravitational junctions ($n\geq 2$) correspond to $n-1$ strings following the non-linear Nambu-Goto equations and coupled to each other with Monge-Amp\'{e}re like terms. Furthermore, for $n>2$, there are non-trivial solutions of the junction conditions in the tensionless limit even in the absence of rigid parameters. This remarkably implies that matter like vibrations can originate from pure gravity. In \cite{Chakraborty:2025jtj}, it was argued that the stringy degrees of freedom of the $n$-way gravitational junctions can be interpreted in terms of combinations of wavepackets on $n-1$ wires which undergo perfect reflection at the interface in a manner similar to the current paper, although in a less rigorous way. Particularly, the role of the rigid parameters have not been analyzed and this requires a global analysis using resummation of the perturbative expansion as has been achieved in the present work. Furthermore, it would be fascinating to understand the holographic interpretation of the tensionless limit of the multiway junction explicitly.

Clearly, the study of the multiway gravitational junctions is incomplete without considering different stationary background geometries (different BTZ geometries) on both sides of the junction, and their distinct deformations in the form of locally AdS$_3$  Ba\~nados spacetimes \cite{Banados:1998gg} corresponding to the deformation of the left/right moving wavepackets encoding the stringy modes of the junction. These should lead to non-trivial $n\rightarrow n$ scattering, generalizing the results of \cite{Bachas:2020yxv,Chakraborty:2025dmc}. Particularly, it would be interesting to understand how the steady state heat current found in interfaces between two wires at different temperatures \cite{Bernard:2016nci,Bachas:2021tnp} generalize to multi-way interfaces, and it would be fascinating to understand non-trivial energy scattering in the background of the steady state currents generalizing the quantum maps studied in \cite{Chakraborty:2025dmc} which realize tunable energy transmitters. In this context, it would also be important to examine the quantum null energy condition and compatibility with quantum thermodynamics following \cite{Kibe:2021qjy,Banerjee:2022dgv,Kibe:2024icu}. Furthermore, it would be fascinating to understand if multi-point stress tensor correlation functions across the interface obtained holographically can be reproduced from CFT methods in general large $c$ CFTs.

Understanding to what extent the stringy dynamics of the multiway junctions can be reconstructed from subregions in the dual field theory and how this reconstruction can be formulated in the framework of quantum error correction \cite{harlow2018tasi,Jahn:2021uqr,Chen:2021lnq,Kibe:2021gtw} is also a problem of fundamental importance. Particularly as multiway gravitational junctions in three dimensional locally AdS$_3$ spacetimes realize tunable quantum processors, the reconstruction of the corresponding dual spacetimes with dynamical junctions would provide a rich source of new understanding of the holographic duality.

Finally, we should be able to obtain stringy degrees of freedom in four and higher dimensional gravity by embedding the multi-way three dimensional junctions as discussed in \cite{Chakraborty:2025jtj}. It would be interesting to understand the holographic reconstruction of these stringy modes in higher dimensional gravity.

\begin{acknowledgments}
We thank Evangelos Afxonidi and Costas Bachas for very valuable discussions. The research of AB is Greece was partially supported by the European MSCA grant HORIZONMSCA-2022-PF-01-01 and by the H.F.R.I call “Basic research Financing (Horizontal support of all Sciences)” under the National Recovery and Resilience Plan “Greece 2.0” funded by the European Union– NextGenerationEU (H.F.R.I. Project Number: 15384). AB also acknowledges the support from the Ulam Postdoctoral Fellowship from the  Polish National Agency for Academic Exchange (NAWA). The research of AM has been supported by the FONDECYT regular grant no. 1240955 of La Agencia Nacional de Investigaci\'{o}n y Desarrollo (ANID), Chile. AM gratefully acknowledges the hospitality of LPENS, where a part of this work was carried out during his tenure as a CNRS invited professor. TK and GP gratefully acknowledge the hospitality of IFIS, PUCV where a large part of this work was carried out during their visit.
\end{acknowledgments}
\appendix

\section{Calculation of the entanglement entropy}\label{App:EE}
We outline the method used to calculate entanglement entropy \eqref{Eq:EEsubtr} in two-dimensional holographic interface CFTs, following the technique developed in \cite{Kibe:2021qjy,Banerjee:2022dgv,Kibe:2024icu}.

We wish to compute the holographic entanglement entropy for a space-like boundary interval as the length of the bulk geodesic anchored at the end-points of the boundary interval \cite{RT,HRT}. The bulk spacetime consists of two copies of a locally $AdS_3$ spacetime glued along a junction that is the worldsheet of a tensile string.
We focus on the case where the left and right spacetimes are both BTZ black holes with the metric
\begin{equation}
    ds^2 = \frac{dz^2}{\frac{z^2}{\ell^2}-4 \mu^2\ell^2}-\lb\frac{z^2}{\ell^2}-4\mu^2\ell^2\rb dt^2+\frac{z^2}{\ell^2} dx^2,
    \label{Eq:metricFG}
\end{equation}
in Fefferman-Graham (FG) gauge, where $\ell$ is the AdS curvature scale.  It is useful to transform this metric to Eddington-Finkelstein (EF) gauge in order to use the uniformization maps from \cite{Kibe:2021qjy,Banerjee:2022dgv,Kibe:2024icu}. The transformation from EF gauge, with metric
\begin{equation}
    ds^2 = -\lb\frac{z^2}{l^2}-4\mu^2l^2\rb du^2+\frac{z^2}{l^2} dx^2 +2 du dz,
    \label{Eq:metricEF}
\end{equation}
to FG gauge is
\begin{equation}\label{Eq:EF-FG}
    u = t-\frac{1}{4\mu} \log\left(\frac{z+2\mu \ell^2}{z-2\mu \ell^2 }\right),
\end{equation}
where the $z$ and $x$ coordinates are unchanged. The above solution for $u$ is real only outside the horizon, that is, when $z>2\mu \ell^2$. The metric \eqref{Eq:metricEF} can be transformed (uniformized) to the vacuum metric
\begin{equation}\label{Eq:metricVac}
    ds^2 = - \frac{Z^2}{\ell^2} dU^2+\frac{Z^2}{\ell^2} dX^2 +2 dU dZ,
\end{equation}
where the vacuum coordinates are denoted by capital letters. The uniformization map that achieves this transformation is \cite{Kibe:2021qjy,Banerjee:2022dgv,Kibe:2024icu}
\begin{align}       \label{eq-uniformization-map}
    &Z= e^{-2\mu u}(z-2\mu \ell^2),\nonumber\\
    &U = \frac{e^{2\mu u}(z \cosh(2\mu x)-2\mu \ell^2 )}{2\mu(z-2\mu \ell^2)},\\
    &X = \frac{e^{2\mu u} z \sinh(2\mu x)}{2\mu(z-2\mu \ell^2)}\nonumber.
\end{align}
We can transform from the FG metric \eqref{Eq:metricFG} to the vacuum metric \eqref{Eq:metricVac} using the uniformization map \eqref{eq-uniformization-map} and the transformation \eqref{Eq:EF-FG}.

\subsection{Computing geodesic lengths using the uniformization map}\label{Sec:method}
\begin{figure}
    \centering
    \includegraphics[scale=0.4]{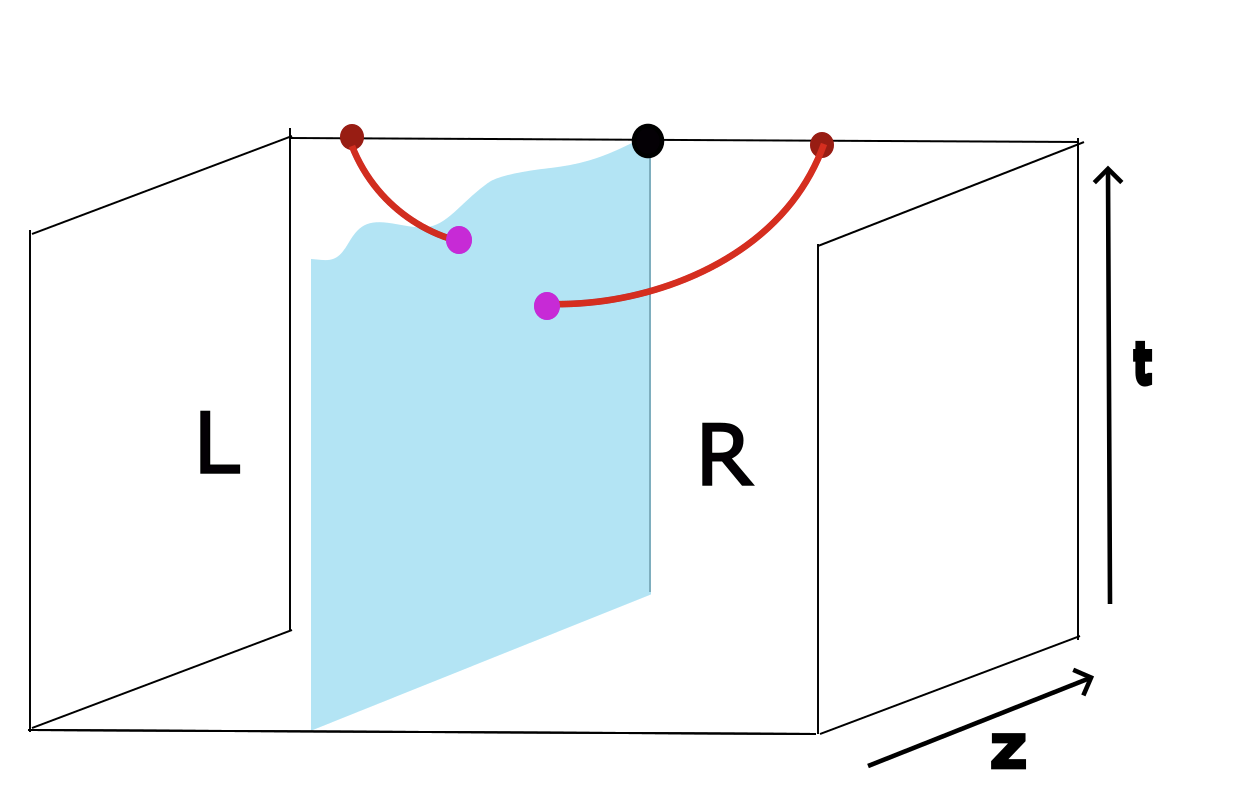}
    \caption{The two spacetimes L and R are glued along the worldsheet in blue. $t$ is the boundary time and $z$ is the radial coordinate. A  geodesic (solid red) connecting the two end-points (red circles) of the boundary interval is shown. The geodesic intersects the worldsheet at the magenta points. The left and right magenta points have the same worldsheet coordinates $\tau_*, \sigma_*$ but have different images in the left and right spacetimes. The black circle denotes $x=0$, where the worldsheet intersects the boundary. }
    \label{Fig:method}
\end{figure}

We wish to compute the entanglement entropy of a spacelike interval with end-points in the left and right spacetime as shown in Fig.~\ref{Fig:method}.
Let the end-points be $(t_{1,2},x_{1,2},z_{1,2})$, where  $z_{1,2} \to \infty$ at the boundary and $x_1<0$ and $x_2>0$. Note that these are the original (discontinuous) coordinates and not the continuous $\tilde{x}^\pm$ coordinates obtained after a half-sided conformal transformation. For the perturbative solutions to be valid we will assume that the length of the boundary interval $l$ is much smaller than the thermal length scale $\mu l \ll 1$, such that the geodesic doesn't penetrate behind the horizon on either side of the worldsheet. The geodesic intersects the worldsheet at a point parametrized by the worldsheet coordinates $(\tau_*,\sigma_*)$. The images of this point in the left and right spacetimes are given by 
\begin{align}
    &t_{L,R}(\tau_*,\sigma_*) = \tau_*\mp \tau_d(\tau_*,\sigma_*), \quad z_{L,R}(\tau_*,\sigma_*)=\sigma_* \mp \sigma_d(\tau_*,\sigma_*), \nonumber\\ & x_{L,R}(\tau_*,\sigma_*) = x_s(\tau_*,\sigma_*) \mp x_d(\tau_*,\sigma_*),
    \label{Eq:wstoBTZ}
\end{align}
where $\tau_d,\sigma_d,x_s,x_d$ are given in \eqref{Eq:wstoBTZ2}.

We regularize the calculation by imposing the cutoff on the radial worldsheet coordinate $\sigma$ as $\sigma_c=\ell^2 \delta_c^{-1}$. The boundary end-points for our interval of interest are then defined as
\begin{align}
    &t_{1,2} = t_{1,2}^0\mp \tau_d\lb t_{1,2}^0,\frac{\ell^2}{\delta_c}\rb,\\
    &z_{1,2}= \frac{\ell^2}{\delta_c}\pm\sigma_d\lb t_{1,2}^0,\frac{\ell^2}{\delta_c}\rb,\\
    & x_{1,2}= x_{L_b,R_b} + \mathcal{O}(\delta_c),
\end{align}
where $t_{1,2}^0$ is the time coordinate of the left and right end-point of the interval when $\epsilon\to0$.

An overview of the method to obtain the geodesic in this setup is:
\begin{enumerate}
    \item Solve for the geodesic tangent $\dot{x}^{\mu} = (\dot{t},\dot{z},\dot{x})$ in the left and right geometries separately in terms of any two quantities conserved along a geodesic in the respective geometries.
    \item Compute these conserved quantities using uniformisation maps.
    \item Find the discontinuity at the worldsheet junction in the tangent to the geodesic ($\delta\dot{x}^{\mu}$)
    \item Combine the above steps to solve for the intersection point $(\tau_*,\sigma_*)$.
\end{enumerate}

We provide details for these four steps below.
\subsubsection{Step 1: Geodesic tangents in terms of conserved quantities}\label{ssec:step1}
In each of the left and right spacetimes we have the two conserved quantities
\begin{align}
   e_{L,R}&= (1,0,0)\cdot g^{(L,R)}\cdot(\dot{t}_{L,R},\dot{z}_{L,R},\dot{x}_{L,R}),\\
   j_{L,R}&=(0,0,1)\cdot g^{(L,R)}\cdot(\dot{t}_{L,R},\dot{z}_{L,R},\dot{x}_{L,R}),
   \label{eq:conservedqt}
\end{align}
where $g^{(L,R)}$ are the metrics on the left and right and $(\dot{t}_{L,R},\dot{z}_{L,R},\dot{x}_{L,R})$ are the geodesic tangents in the left and right spacetimes, respectively.
The conserved quantities $e,j$ correspond to time and spatial translation symmetries respectively. Note that these quantities are conserved separately in the left and right spacetimes but are not globally conserved. Using these quantities and the normalization of the geodesic as
\begin{equation}
  (\dot{t}_{L,R},\dot{z}_{L,R},\dot{x}_{L,R})  \cdot(\dot{t}_{L,R},\dot{z}_{L,R},\dot{x}_{L,R})=1,
\end{equation}
we can solve for the tangent $(\dot{t}_{L,R},\dot{z}_{L,R},\dot{x}_{L,R})$ in terms of the conserved quantities as
\begin{align}
    &\dot{t}_{L,R} = \frac{e_{L,R} \ell^2}{4\mu^2 \ell^4-z_{L,R}^2},\\
    &\dot{x}_{L,R} = \frac{j_{L,R} \ell^2}{z_{L,R}^2},\\
    &\dot{z}_{L,R} = \mp\frac{\sqrt{e_{L,R}^2\ell^2z_{L,R}^2+(j_{L,R}^2\ell^2-z_{L,R}^2)(4\mu^2\ell^4-z_{L,R}^2)}}{\ell z_{L,R}},
\end{align}
where $z_{L,R}$ is defined at the intersection point $(\tau_*,\sigma_*)$ using \eqref{Eq:wstoBTZ}.
Thus, we have the tangents at the intersection point in terms of $(\tau_*,\sigma_*)$ and the conserved quantities $(e_{L,R},j_{L,R})$. The $\pm$ sign above must be chosen depending on if the geodesic intersects the $x=0$ hypersurface at a point with increasing or decreasing $z$ coordinate along the geodesic. 

\subsubsection{Step 2: Computing conserved quantities using the uniformization map}\label{ssec:step2}
The conserved quantities $(e_{L,R},j_{L,R})$ defined in the previous subsection can now be computed in vacuum coordinates using the uniformization map. The conserved quantities can be expressed in terms of the end-points of the boundary interval and the intersection points $(\tau_*,\sigma_*)$, and series expanded in $\epsilon$. 

A spacelike geodesic between two arbitrary points $(Z_{1,2},U_{1,2},Y_{1,2})$ in vacuum coordinates, in terms of the length $\xi$ along the geodesic, is
\begin{equation}
    \lb\frac{\ell^2}{Z},U,Y\rb = \lb \frac{\sech \lb\frac{\xi}{\ell}\rb}{\sqrt{\Lambda^2-E^2}},\, U_0 - \frac{\sech \lb\frac{\xi}{\ell}\rb}{\sqrt{\Lambda^2-E^2}} - \frac{E}{\Lambda^2-E^2}\tanh \lb\frac{\xi}{\ell}\rb,\, Y_0 + \frac{\Lambda}{\Lambda^2-E^2}\tanh \lb\frac{\xi}{\ell}\rb \rb, 
    \label{Eq:geo-vacuum}
\end{equation}
where the integration constants $U_0,Y_0, \Lambda, E$ are determined in terms of the end-points $(Z_{1,2},U_{1,2},Y_{1,2})$ \cite{Kibe:2024icu}. Note that $\xi \in (-\infty, \infty)$.

We obtain the left and right geodesic segments as follows. The boundary end-points are plugged into the uniformization map to obtain their images in vacuum coordinates. The image of the intersection points $z_{L,R}(\tau_*,\sigma_*),t_{L,R}(\tau_*,\sigma_*),x_{L,R}(\tau_*,\sigma_*)$ under the uniformization map is also obtained. Thus, using \eqref{Eq:geo-vacuum} we have the left (right) geodesic in terms of the left (right) boundary end-point and the intersection point $\tau_*,\sigma_*$.

The conserved quantities \eqref{eq:conservedqt} are then readily calculated by transforming the Killing vectors $(1,0,0)$, $(0,0,1)$, the metric, and the tangents $(\dot{t}_{L,R},\dot{z}_{L,R},\dot{x}_{L,R})$ to vacuum coordinates. The geodesic \eqref{Eq:geo-vacuum} is used to obtain the tangents. Thus we have the conserved quantities $e_{L,R}$ and $j_{L,R}$ in terms of the boundary end-points and the intersection point.

\subsubsection{Step 3: Discontinuity in the tangents}\label{ssec:step3}
The equation for the discontinuity can be obtained by varying the total geodesic length and demanding that the boundary terms on the worldsheet vanish. This condition is \cite{Papadopoulos:2023kyd}
\begin{equation}
    \frac{t_{a,L}^\nu \dot{x}^\mu_L g^{(L)}_{\mu \nu}}{\sqrt{g^{(L)}_{\mu \nu}\dot{x}^\mu_L\dot{x}^\nu_L}}= \frac{t_{a,R}^\nu \dot{x}^\mu_R g^{(R)}_{\mu \nu}}{\sqrt{g^{(R)}_{\mu \nu}\dot{x}^\mu_R\dot{x}^\nu_R}},  
\end{equation}
where both the left and right sides are evaluated at the point of intersection between the geodesic and the worldsheet. Above, $L,R$ indicate the left and right spacetimes and $t_a$ are tangents to the worldsheet defined as
\begin{equation}
    t_a^\nu = \frac{\partial x^\nu}{\partial \xi^a}, \quad \xi^a = (\tau,\sigma).
\end{equation}
To compute the entanglement entropy we will consider spacelike geodesics and normalize the geodesic tangents. Thus, the equation for the discontinuity becomes
\begin{equation}
    t_{a,L}^\nu \dot{x}^\mu_L g^{(L)}_{\mu \nu}= t_{a,R}^\nu \dot{x}^\mu_R g^{(R)}_{\mu \nu}.
\end{equation}

\subsubsection{Step 4: Equations for the intersection point}\label{ssec:step4}
Combining steps $1-3$ we obtain two equations for the two coordinates of the intersection point $(\tau_*, \sigma_*)$. The explicit equations are cumbersome and we do not provide them here. 

In order to obtain a solution, we series expand these equations and solve them order by order in $\epsilon$. As stated before, we also assume that the length of the boundary interval $l$ is much smaller than the thermal length, i.e., $\mu l \ll1$, and series expand in $\mu l$. This ensures that the perturbative solution for the gravitational junction is valid. The equations take a much simpler form and can be readily solved when expanded in these two parameters. Note that the time coordinates of the end-points of the boundary interval are arbitrary and are not assumed to be small compared to the thermal length.

\subsubsection{Geodesic length}
The advantage of using the uniformization map is that the proper lengths of the two geodesic segments in the left and right geometry can be readily calculated. Once we map the endpoints of each of these segments to the Poincar\'{e} patch via uniformization maps, the length of the geodetic arcs can be computed as follows. If $(Z_1,U_1,Y_1)$ and $(Z_2,U_2,Y_2)$ are the endpoints of a geodesic in the Poincar\'{e} patch, its proper length $L_{\rm geo}$ in units of the AdS scale $\ell$ is given by 
\begin{equation}\label{eq:qnechom-Lgeo}
    L_{\rm geo} = \ell \log(\chi + \sqrt{\chi^2 -1}), \quad \chi = \frac{\lb\frac{\ell^4}{Z_1^2} + \frac{\ell^4}{Z_2^2} - \lb U_1 +\frac{\ell^2}{Z_1} - U_2 -\frac{\ell^2}{Z_2}\rb^2 + (Y_1 - Y_2)^2\rb}{2 \frac{\ell^4}{Z_1 Z_2}}.
\end{equation}
The entanglement entropy of the boundary interval is then obtained via the HRRT prescription, which states that
\begin{equation}
    S_{\rm EE} = \frac{c}{6} \frac{\left(L_{\rm geo}^{L}+L_{\rm geo}^{R}\right)}{\ell},
\end{equation}
with $L_{\rm geo}^{L}$ and $L_{\rm geo}^{R}$ denoting the proper lengths of the geodetic segments in the left and right geometry, respectively. The lengths of $L_{\rm geo}^{L}$, $L_{\rm geo}^{R}$ are regulated by considering that the end-points of the entangling interval lie at the regulated boundary. Clearly, $L_{\rm geo}^{L}$ and $L_{\rm geo}^{R}$ can be easily determined via the corresponding uniformization maps once the intersection point of the geodesic with the worldsheet $(\tau_*,\sigma_*)$ is determined.

\section{Strong sub-additivity}\label{App:SSA}
In this section, we present  details of the strong sub-additivity (SSA) calculation discussed in Sec.~\ref{Sec:SSA}.

Consider the choice of boundary sub-regions shown in Fig.~\ref{Fig:SSA1}. The SSA is
\begin{equation}
S(A) + S(B)-S(A\cup B)-S(A \cap B)\geq0.
\end{equation}
For the vacuum state the proper lengths of the regions are chosen to be 
$|A \cup B| = r_2$, $|A \cap B| = r_1 $ and $|A| = |B|=\sqrt{r_2 r_1} $. Note that $r_2>r_1$. This choice was first considered by \cite{Casini:2006es} to prove the c-theorem and ensures that the SSA is saturated for the CFT vacuum.

Note that we will obtain a non-trivial SSA only if the defect worldline intersects at least one of the four subregions shown in Fig.~\ref{Fig:SSA1}. It is easy to see that the number of regions intersected by the defect is either zero, two or four. The cases with four (two) regions intersected by the defect, are shown in Fig.~\ref{Fig:SSA1} (a) and (b). Furthermore, in order to obtain constraints on the defect we must ensure that the SSA is saturated at leading order in the $\epsilon $ and $\mu l$ expansions. The choice of regions shown in Fig.~\ref{Fig:SSA1} ensures that this is the case at order $\epsilon^0 (\mu l)^0$. However, our setup involves gluing two BTZ black holes, which are dual to thermal states in the CFT. We can obtain the correct sub-region configuration, which saturates SSA for $\epsilon\to0$ to all orders in $\mu l$, by conformally transforming the above regions using a transformation that maps the vacuum to a thermal state. Since the SSA doesn't change under a conformal transformation, it is saturated for the thermal state with these transformed subregions.

Let us choose the end-points of the intervals, in the absence of the defect $\epsilon\to0$, by transforming the vacuum interval as follows \footnote{Note that these are the end-points in the discontinuous coordinates $t,x$ and not in the continuous coordinates $\tilde{t},\tilde{x}$ one obtains after a conformal transformation on the right.}
\begin{align}
    &x^\pm_{L_A}=\frac{\log \left(\frac{\mu(r_2-r_1)}{2}\mp\mu\left(\frac{r_1}{2}+x_0\right)+\mu t_0\right)}{2 \mu },\quad x^\pm_{R_A}=\frac{\log \left(\pm\mu  \left(\frac{r_2}{2}-x_0\right)+\mu  t_0\right)}{2 \mu } \nonumber\\
    &x^\pm_{L_B}=\frac{\log \left(\mu t_0\mp\mu  \left(\frac{r_2}{2}+x_0\right)\right)}{2 \mu }, \quad x^\pm_{R_B}=\frac{\log \left(\frac{\mu  (r_2-r_1)}{2}\pm\mu  \left(\frac{r_1}{2}-x_0\right)+ \mu t_0\right)}{2 \mu },\label{Eq:ssaendpts}\\
    &x^\pm_{L_{A\cup B}}=x_{L_B}^\pm, \quad x^\pm_{R_{A\cup B}}=x_{R_A}^\pm, \quad x^\pm_{L_{A\cap B}}=x_{L_A}^\pm, \quad x^\pm_{R_{A\cap B}}=x_{R_B}^\pm,\nonumber
    \label{Eq:ssaendpts}
\end{align}
where $x^\pm= t \pm x$ are the lightcone coordinates in the boundary theory, the subscripts $L,R$ indicate the left and right end-points of the corresponding intervals $A,B, A\cup B$ or $A\cap B$. The defect is located at $x=0\, ,t=\frac{\log(\mu t_0)}{2\mu}$. The $x_0$ parameter is the distance between the defect and the centre of $A\cup B$ in Fig.~\ref{Fig:SSA1}, i.e., the x-coordinate of the centre is $-x_0$.

In the presence of the defect the end-points above are perturbatively corrected based on the time re-parametrization at the defect. However, note that we must be careful to ensure that the perturbatively corrected end-points have the correct domains of dependences. In Fig.~\ref{Fig:SSA1} it is essential to ensure that the equal time intervals labelled $A\cup B$ and $A\cap B$ have the same domains of dependences as the union (intersection) of the domains of $A$ and $B$. This feature is retained when we perform a conformal transformation and map to \eqref{Eq:ssaendpts}. This consistency of the causal diamonds must also be ensured in the choice of end-points in the presence of the defect, i.e, when we include $\epsilon$ corrections. We ensure this as follows. Shoot two null rays from every undeformed end-point in \eqref{Eq:ssaendpts} and find the time at which they intersect the defect worldline ($x=0$). Use this intersection time in the time-reparametrization and find the $\epsilon-$corrected intersection time. The two $\epsilon$-corrected null rays are completely specified using this procedure and this uniquely determines how the end-point should be shifted in the presence of the defect. It is easy to see that this transformation of null rays will ensure that the domains of dependences are consistent. For example, this ensures that the left (right) end-points of $A$ and $B$ remain null separated to the desired order in the $\epsilon$ expansion. Note that making this choice of shifted end-points requires us to also perturbatively shift the spatial coordinates compared to \eqref{Eq:ssaendpts}.

Using these end-points and the method described in Appendix.~\ref{App:EE}, we can easily compute the SSA. Note that we assume that $\mu r_{1,2} \ll1$, $\mu x_0 \ll1$ and series expand in these dimensionless parameters.  We describe the results for the SSA in the two cases shown in Fig.~\ref{Fig:SSA1} in the following subsections.

\subsection{Case (a): Defect intersects all four intervals}
We first consider the configuration shown in Fig.~\ref{Fig:SSA1} (a). This corresponds to imposing the following condition on the end-points \eqref{Eq:ssaendpts}
\begin{equation}
    \frac{r_1}{2}\pm x_0>0, \quad \frac{r_2}{2}\pm x_0 > 0.
    \label{Eq:r12casea}
\end{equation}
In this case the SSA is
\begin{equation}
    \epsilon\Bigg((\log g_0)\lb\sqrt{1-2\frac{x_0}{r_2}}-\sqrt{1-2\frac{x_0}{r_1}}\rb\lb\sqrt{1+2\frac{x_0}{r_1}}-\sqrt{1+2\frac{x_0}{r_2}}\rb+\mathcal{O}(\mu r_2, \mu r_1)\Bigg) \geq0,
\end{equation}
where the term multiplying $\log g_0$ is manifestly positive since $r_2>r_1$, and the SSA is therefore satisfied.

We can also examine the SSA at higher orders. However, the higher order terms in SSA can place bounds on the parameter space only if the leading order pieces vanish. The only way to achieve this without setting $\lambda=0$ is to choose $x_0=0$, that is, for an interval that is symmetric about the defect. In this case, we have checked that the SSA is saturated to $\mathcal{O}(\epsilon^2 \mu r_{1,2})$. Note that this is the order at which the Nambu-Goto mode $A_0$ contributes to the entanglement entropies. Note also that the above expression is obtained without setting $\gamma_h=0$. Thus, the SSA is satisfied perturbatively even in the presence of the rigid parameters $\alpha_h$, $\gamma_h$ and the Nambu-Goto mode $A_0$.

\subsection{Case (b): Defect intersects two intervals}
We now consider the configuration shown in Fig.~\ref{Fig:SSA1} (b), which corresponds to imposing
\begin{equation}
    \frac{r_1}{2}- x_0<0, \quad \frac{r_1}{2}+x_0>0, \quad \frac{r_2}{2}\pm x_0 > 0
    \label{Eq:r12caseb}.
\end{equation}
The SSA in this case is
\begin{equation}
    \epsilon\left((\log g_0) \left(\sqrt{\lb1-2\frac{x_0}{r_2}\rb\lb1+2\frac{x_0}{r_1}\rb}-\sqrt{\lb1-2\frac{x_0}{r_2}\rb\lb1+2\frac{x_0}{r_2}\rb}\right)\right) \geq0.
\end{equation}
Similar to case (a) we find that the SSA is satisfied, which can be easily seen as
\begin{multline}
     \sqrt{\lb1-2\frac{x_0}{r_2}\rb\lb1+2\frac{x_0}{r_2}\rb}-\sqrt{\lb1-2\frac{x_0}{r_2}\rb\lb1+2\frac{x_0}{r_1}\rb}<\\ \quad \sqrt{\lb1-2\frac{x_0}{r_2}\rb\lb1+2\frac{x_0}{r_2}\rb}-\sqrt{\lb1-2\frac{x_0}{r_2}\rb\lb1+2\frac{x_0}{r_2}\rb}=0,
\end{multline}
since $r_2>r_1$.

As we saw for case (a), the subleading pieces of the SSA become relevant only if  $x_0=0$. However, \eqref{Eq:r12caseb} implies that $x_0>\frac{r_1}{2}>0$. Thus, we cannot set $x_0=0$ in case (b), and the higher order terms in the SSA do not impose a bound on the parameter space. The other possibility for the SSA to be saturated at this order is when $x_0=\frac{r_2}{2}$, i.e, when the defect is exactly at the right end-point of $A$ and $A\cup B$. In this case, perturbatively, the geodesic is entirely in the left geometry and we expect the SSA to be saturated to all orders. In ICFTs, entanglement entropies of intervals with one end-point at the defect can have modified coefficients for the leading log divergent pieces; this is thought of as a new effective central charge. However, this effect will not be seen by the perturbative setup studied here, and we postpone an analysis of the effective central charge to a future publication.

\bibliographystyle{JHEP}
\bibliography{references}

\end{document}